\documentclass[prd,twocolumn,tightenlines,showpacs,nofootinbib,amsfonts,amssymb,amsmath]{revtex4-1}
\usepackage{graphicx}
\begin{document}
\newcommand{\etal}{{\it et al.}}
\newcommand{\bx}{{\bf x}}
\newcommand{\bn}{{\bf n}}
\newcommand{\bk}{{\bf k}}
\newcommand{\dd}{{\rm d}}
\newcommand{\dslash}{D\!\!\!\!/}
\def\ga{\mathrel{\raise.3ex\hbox{$>$\kern-.75em\lower1ex\hbox{$\sim$}}}}
\def\la{\mathrel{\raise.3ex\hbox{$<$\kern-.75em\lower1ex\hbox{$\sim$}}}}
\def\beq{\begin{equation}}
\def\eeq{\end{equation}}

\leftline{UMN--TH--3119/12}

\vskip-2cm
\title{The anisotropic power spectrum and bispectrum in the $f \left( \varphi \right) F^2$ mechanism}

\author{Nicola Bartolo$^{1,2}$, Sabino Matarrese$^{1,2}$, Marco Peloso$^{1,3}$, 
Angelo Ricciardone$^{1,2}$}
\affiliation{
${^1}$ Dipartimento di Fisica e Astronomia ÒG. GalileiÓ, \\
Universit\`a degli Studi di Padova, I-35131 Padova (Italy) \\
${^2}$ INFN, Sezione di Padova, I-35131 Padova  (Italy)\\
${^3}$ School of Physics and Astronomy,
University of Minnesota, Minneapolis, 55455 (USA)\\
}
\vspace*{2cm}

\begin{abstract}
A suitable coupling of the inflaton $\varphi$ to a vector kinetic term $F^2$ 
gives  frozen and scale invariant vector perturbations. We compute the
cosmological perturbations $\zeta$ that result from such coupling by taking into
account the classical vector field that unavoidably gets generated at
large scales during inflation.  This generically results in a too anisotropic
power spectrum of $\zeta$. Specifically, the anisotropy exceeds the 
 $1\%$ level ($10 \%$ level)  if inflation lasted  $\sim 5$ e-folds ($\sim 50$ e-folds) 
more than the minimal amount required to produce the CMB modes. This conclusion applies, 
among others, to the application of this mechanism for  magnetogenesis, for  anisotropic inflation,
and for the generation of anisotropic perturbations  at the end of inflation through a waterfall 
field coupled to the vector (in this case, the unavoidable contribution that we obtain  is 
effective all throughout inflation,  and it  is  independent of the waterfall field). 
 For a tuned duration  of inflation, a $1\%$ ($10 \%$)  anisotropy in the power spectrum 
  corresponds to an anisotropic bispectrum which is enhanced  like the local one in the squeezed limit, 
 and with an effective local $f_{\rm NL} \sim 3 \,  (\sim 30)$.
More in general, a significant anisotropy of the perturbations may be a natural outcome of all models that sustain 
higher  than $0$ spin fields during inflation.

\end{abstract}
 \date{September 2012}
 \maketitle

\section{Introduction}

In this work we compute the power spectrum and the bispectrum of the cosmological curvature perturbations $\zeta$ under the assumptions that 
\begin{enumerate}
\item The dominant contribution to $\zeta$ is provided by a slowly rolling inflaton field

\item The inflaton is coupled to the kinetic term of a vector field so to produce a nearly scale invariant and frozen spectrum of vector perturbations at large scales.
\end{enumerate}

These assumptions are realized in   models of magnetogenesis and of  statistical anisotropy of the cosmological perturbations. Despite the phenomenology of these models has been heavily studied in the literature, we obtain novel and  general results, since,  for the first time to our knowledge, we simultaneously take into account both the facts that (i) a strong contribution to the anisotropy (both in the power spectrum and in the bispectrum of $\zeta$) results from the same coupling characterized in 2., and (ii) the inflationary expansion that took place before the Cosmic Microwave Background (CMB)  modes left the horizon unavoidably results in a classical background   vector field that is homogeneous from the point of view of the CMB modes, but breaks isotropy.

Interest for models  that can produce vector fields during inflation has been generated by the inference of intergalactic magnetic fields and by claims of broken statistical invariance of the CMB modes.  Intergalactic magnetic fields have been inferred by an apparent lack of GeV scale  $\gamma-$rays coming from blazars that produce TeV scale $\gamma-$rays; in standard models, 
part of the higher energy  $\gamma-$rays should be converted in lower energy secondaries (which then generate the lower energy $\gamma-$rays) by their interaction with the inter galactic medium. The non-observation of the  GeV scale  $\gamma-$rays   has been explained as  intergalactic magnetic fields that deflect the secondaries \cite{Neronov:1900zz}. 

Broken statistical isotropy of the CMB perturbations has instead been found in the studies \cite{Groeneboom:2008fz,Hanson:2009gu,Groeneboom:2009cb} of the WMAP data. While the overall WMAP results  \cite{Komatsu:2010fb} strongly support the inflationary paradigm, the above studies have shown that  the statistics of the WMAP anisotropies does not possess full rotational  invariance. Specifically, under the parametrization  \cite{Ackerman:2007nb} 
\begin{equation}
P_\zeta \left( \vec{k} \right) = P \left( k \right) \left[ 1 + g_* \, \cos^2 \theta_{ {\hat k} ,\, {\hat V} }  \right]
\label{acw}
\end{equation}
(which can be thought of as an expansion series of the power spectrum in the limit of small anisotropy, truncated at the quadruple term) the WMAP data give $g_* = 0.29 \pm 0.031$ \cite{Groeneboom:2009cb}.  The ``privileged'' direction $\hat{V}$ lies very close to the ecliptic poles. This strongly suggests a systematical origin of the effect, and it has been shown in  \cite{Hanson:2009gu,Hanson:2010gu} that the instrument beam asymmetry can account for it.
Fortunately, the Planck satellite will soon provide an independent test of this with an expected sensitivity to a quadrupolar anisotropy in the power spectrum as small as $0.5\%$ at 1$\sigma$ \cite{Pullen:2007tu,MA}. 
On different scales (and marginalizing over the preferred direction $\hat V$) Large-Scale Structure data analysis constrain $-0.41 < g_* < 0.38$ at $95 \%$ C.L. \cite{Pullen:2010zy} (the amplitude of the anisotropy may in general  be scale dependent~\cite{Ackerman:2007nb}).  

Broken rotational invariance could be the result of anisotropic inflation \cite{Gumrukcuoglu:2006xj}. It is however nontrivial to realize this, since anisotropic spaces typically rapidly isotropize in presence of a cosmological constant \cite{Wald:1983ky}.~\footnote{See \cite{Maleknejad:2012as} for the extension of the study of  \cite{Wald:1983ky} to slow roll inflation} Vector fields may in principle support the anisotropy. In this case,  the problem of preserving the anisotropy translates into contrasting the quick decrease of the vector energy that takes place for a minimal  ${\cal L}_A = - F^2/4$. To our knowledge, four distinct classes of models have been constructed to achieve this; the first three of them are characterized by  (i) a vector potential $V \left( A^2 \right)$ \cite{Ford:1989me}, (ii) a fixed vector vev due to a lagrange multiplier \cite{Ackerman:2007nb}, and (iii) a vector coupling $A^2 R$ to the scalar curvature $R$  \cite{Golovnev:2008cf,Kanno:2008gn}. This last mechanism was originally employed for magnetogenesis in \cite{Turner:1987bw}. These three proposals break the U(1) symmetry of the minimal action, and lead to an additional degree of freedom, the longitudinal vector polarization, that in all of these models turns out to be a ghost  \cite{hcp}. 

The fourth class is instead U(1) invariant  and free of ghost instabilities. It is characterized by a  function of a scalar inflaton $\varphi$ multiplying the vector kinetic term, 
\begin{equation}
{\cal L} = - \frac{I^2 \left( \varphi \right)}{4} F_{\mu \nu} F^{\mu \nu}
\label{L-ratra}
\end{equation}
A suitably chosen evolution for $\langle I \rangle$ during inflation results in a (nearly) constant vector energy density, and therefore in a prolonged anisotropic expansion  \cite{Watanabe:2009ct}.~\footnote{See \cite{aniso-fAA} for models of anisotropic inflation  that employ the idea of  \cite{Watanabe:2009ct}. Also, an interesting model of vector curvaton \cite{Dimopoulos:2006ms}  employing a varying mass $m$ and kinetic function $I$ has been proposed in 
\cite{Dimopoulos:2009am} and studied in \cite{Afm-study,Namba:2012gg}. In particular, ref. \cite{Namba:2012gg} demonstrated that  treating $I$ and $m$ as functions of a quantum inflaton field results in a different phenomenology than just treating them as classical external functions.} Also this mechanism was originally suggested for magnetogenesis  \cite{Ratra:1991bn}  (this application is however problematic \cite{Demozzi:2009fu}, as we discuss below). 
For anisotropic expansion,  a homogeneous vector field pointing along a given direction corresponds to an ``electric'' component, and (\ref{L-ratra})   enjoys an  ``electric'' $\leftrightarrow$ ``magnetic'' duality  under  $I^2 \leftrightarrow \frac{1}{I^2}$ \cite{Giovannini:2009xa}. A constant ``electric'' component is also produced through (\ref{L-ratra}) in the mechanism of  \cite{Yokoyama:2008xw}, in which the vector field  is coupled to the  waterfall field $\chi$ of hybrid inflation through a $\chi^2 A^2$ interaction.  Due to this, the gauge field provides a contribution to the mass of $\chi$,  concurring to determine the moment at which inflation ends, and - thanks to this - contributing to the curvature perturbation. The waterfall field acts as the medium through which the anisotropy in $A_\mu$ is communicated to the inflation; however, as we shall see, the communication already 
occurs through the very same interaction (\ref{L-ratra}) that supports the vector field. This unavoidable effect has not been accounted for neither in  \cite{Yokoyama:2008xw}, nor in the related works  \cite{Karciauskas:2008bc,Karciauskas:2011fp,Emami:2011yi,Lyth:2012br,Lyth:2012vn}. 

The linearized theory of cosmological perturbations in the anisotropic inflationary model of   \cite{Watanabe:2009ct} was worked out in  \cite{Himmetoglu:2009mk,Dulaney:2010sq,Gumrukcuoglu:2010yc,Watanabe:2010fh}. The classical equations of motion of the model admit an attractor solution, \cite{Watanabe:2009ct,Hervik:2011xm} characterized by a non-vanishing ``electric'' component $\vec{E}^{(0)}$. A $10\%$ level anisotropy ($\vert g_* \vert = {\rm O } \left( 0.1 \right)$) is found for an energy $\vert \vec{E}^{(0)} \vert^2 / 2 $ which is about eight orders of magnitude smaller than the inflaton potential \cite{Dulaney:2010sq,Gumrukcuoglu:2010yc,Watanabe:2010fh}. Therefore, the vector energy needs to be highly subdominant not too produce a too strong anisotropy. The work \cite{Barnaby:2012tk}  computed instead the cosmological perturbations in the case in which (\ref{L-ratra}) provides scale invariant ``magnetic'' components of the vector field, as in the magnetogenesis application  \cite{Ratra:1991bn,Martin:2007ue}. Cosmological applications in this context have also been studied in  \cite{ratra-pert}. Studies of the cosmological perturbations in  \cite{Ratra:1991bn,Martin:2007ue} start from the point of view that the statistics of the generated ``magnetic'' field is isotropic, and therefore obtain statistical isotropic results. In shorts, in the magnetogenesis context  \cite{Ratra:1991bn,Martin:2007ue} one does not have the analogous of the attractor solution $\vec{E}^{(0)}$ of the classical equations of motion of the anisotropic inflationary model  \cite{Watanabe:2009ct} (the corresponding $\vec{B}^{(0)}$ vanishes in  \cite{Ratra:1991bn,Martin:2007ue}) and therefore it is simply assumed that $g_* = 0$ in this case.

However, in  \cite{Watanabe:2009ct,Yokoyama:2008xw} (respectively, in \cite{Ratra:1991bn,Martin:2007ue}), the CMB perturbations are affected by a classical ``electric'' field  $\vec{E}_{\rm classical}$ (respectively,  classical ``magnetic'' field  $\vec{B}_{\rm classical}$) which is in general different from the value  given by the classical equations of motion. Indeed, such mechanisms are designed to result in a nearly scale invariant spectrum for the ``electric'' (respectively, ``magnetic'') perturbations. Let us denote by $N_{\rm tot}$ the number of e-folds of inflation. The modes that left the horizon in the first $N_{\rm tot} - N$ e-folds of inflation  add up as a classical background from the point of view of the modes that leave the horizon in the final  $N$ e-folds. This is well appreciated for scalar fields during inflation \cite{Linde:2007fr}. The modes that leave the horizon add up in a ``random walk'' manner to form a classical background that is experienced as  homogeneous by modes of smaller size. A homogeneous classical vector is a field that points in a given direction (that, in a given realization of the model, is determined by the random addition of the super-horizon modes) that breaks isotropy. In the magnetogenesis application, the effects of this  energy  $\langle \vec{B}^2 \rangle / 2$ on the background evolution have been well appreciated 
(see, among others, \cite{Demozzi:2009fu,Kanno:2009ei,Fujita:2012rb}). We point out that this energy is associated with a classical vector field, and in this work we show that this vector imprints a strong anisotropy to the power spectrum and bispectrum of $\zeta$ in all the applications of (\ref{L-ratra}).

Specifically, we show that the natural value of $g_*$ associated to these modes is $\sim 0.1$ (respectively, $\sim 0.01$), if inflation lasted about $50$ e-folds (respectively, about $5$ e-folds) more than the final $\sim 60$ e-folds necessary to generate the CMB modes. Generic models of slow roll inflation are characterized by a much longer duration of inflation, and, therefore, embedding the mechanism (\ref{L-ratra}) in one of these models generically results in too anisotropic perturbations. For a tuned duration of inflation the mechanism becomes extremely predictive since, there is essentially no free parameter in (\ref{L-ratra}). The only relevant quantity is the magnitude of the classical vector field present when the CMB modes left the horizon, and that can be ``traded'' for $g_*$. Therefore, any given value of $g_*$ should be associated with firm predictions for other observables. 

 There are two such predictions that immediately come to mind: the first is a TB and an EB mixing in the CMB data \cite{Watanabe:2010bu}, resulting from the coupling between scalar and tensor modes that, due to the anisotropy, takes place already in the linearized theory \cite{Gumrukcuoglu:2006xj,Pereira:2007yy,Gumrukcuoglu:2007bx,Pitrou:2008gk}. The second is a  directionality dependence in the bispectrum (and, in principle, in the higher point correlation functions), with a clear correlation with the one in the power spectrum.  The bispectrum  resulting from (\ref{L-ratra}) is  computed for the first time in the present work.~\footnote{ For previous works on anisotropic non-gaussianity, see~\cite{Yokoyama:2008xw,Dimopoulos:2008yv,Karciauskas:2008bc,N1,N2,Manureview,Karciauskas:2011fp,Dey:2011mj,N3,Lyth:2012br,Dey:2012qp,Lyth:2012vn}.}
We show that the isotropic  power spectrum and bispectrum of \cite{Barnaby:2012tk}   are in fact the theoretical expectation 
(i.e., the theoretical average over several realizations) for the anisotropic signals that we obtain here, and which are the real quantities that are produced by  any single realization of (\ref{L-ratra}). Quite interestingly, an observable $g_*$ produced from  (\ref{L-ratra}) is associated to an observable bispectrum which is enhanced like the local one in the squeezed limit, and which has a characteristic shape and anisotropy (immediately correlated with the one in the power spectrum).

The paper is organized as follows. In Section \ref{sec:gauge} we study the spectrum of the vector field perturbations obtained from (\ref{L-ratra}), with a particular attention for the functions $I$ that result in scale invariant vector modes. In this section we also further discuss the role of the large-wavelength modes in determining the classical background anisotropy that affects modes of CMB wavelengths. In Section \ref{sec:perturbations} we study how these modes $\delta A$ are coupled to the modes of $\zeta$ through  (\ref{L-ratra}). The power spectrum and bispectrum of $\zeta$ are computed respectively in Sections \ref{section:zz} and \ref{section:zzz}. There, we show explicitly how the sum of the long-wavelength modes adds up with the solution of the classical equations of motion to determine the physical value of $\vec{E}_{\rm classical}$ (or $\vec{B}_{\rm classical}$) observed by the CMB modes. The resulting phenomenology is reviewed in Section  \ref{sec:phenomenology}. In Section \ref{sec:apps} we discuss our results, that generally apply to all the realizations of (\ref{L-ratra}) that give a nearly scale invariant vector field, in the context of anisotropic inflation \cite{Watanabe:2009ct}, of the waterfall mechanism  \cite{Yokoyama:2008xw}, and of magnetogenesis 
\cite{Ratra:1991bn,Martin:2007ue}. A concluding discussion is given in Section \ref{sec:conclusions}.

\section{A scale invariant vector field }
\label{sec:gauge}

Let us consider a  locally U(1) invariant vector field  with lagrangian (\ref{L-ratra}). Ref. \cite{Ratra:1991bn} identified this field with the electromagnetic one, assuming  that $I$ sets to a constant after inflation (in this case, we can simply normalize $I_{\rm end} = 1$). The function $I$ enters in the definition of the electric and magnetic components 
\begin{equation}
E_i = -  \frac{ \langle I \rangle }{a^2} \, A_i' \;\;,\;\;
B_i =   \frac{\langle I \rangle}{a^2} \, \epsilon_{ijk} \partial_j A_k
\label{EB-ratra}
\end{equation}
(we denote by $\langle \dots \rangle$ the vacuum expectation value of a field, or of a function), where prime denotes derivative with respect to conformal time $\tau$, and $a$ is the scale factor of the universe, $d s^2 = a^2 \left( \tau \right) \left( - d \tau^2 + d \vec{x}^2 \right)$.  With the notation (\ref{EB-ratra}), the physical energy density in the vector field assumes the conventional expression $\rho = \frac{\vert \vec{E} \vert^2 + \vert \vec{B} \vert^2}{2}$ at all times. In this work, apart from where we explicitly refer to the magnetogenesis application   \cite{Ratra:1991bn} , we do not necessarily identify the vector field with our photon, but we keep the  ``electromagnetic'' notation (\ref{EB-ratra}) for   convenience. The classical equations of motion obtained from (\ref{L-ratra}) are solved by a homogeneous ``electric'' field $\vec{E}^{(0)} \propto \frac{1}{a^2 \, \langle I \rangle}$. While  the standard case, $I = {\rm const}$, corresponds to $\rho_E \propto a^{-4}$, a constant ``electric'' energy is obtained if $\langle I \rangle \propto a^{-2}$. 

A desired time evolution for $\langle I \rangle$ can be obtained for several functions $I \left( \varphi \right)$, provided they are suitably arranged with the inflaton potential  \cite{Martin:2007ue}. Indeed, 
\begin{equation}
a \propto {\rm exp } \left[ - \int \frac{d \varphi}{\sqrt{2 \epsilon \left( \varphi \right)} M_p} \right] 
\label{a-phi}
 \end{equation}
where we have introduced the slow roll parameter
\begin{equation}
\epsilon \equiv  \frac{M_p^2}{2} \, \left( \frac{V'}{V} \right)^2
\end{equation}
(prime on a function here denotes derivative with respect to its argument) and where a monotonic slow roll inflaton evolution with $\dot{\varphi} < 0$ is assumed. Therefore, a desired behavior $\langle I \rangle = f \left( a \right)$ can be obtained by choosing the functional form of $I$ to coincide with that function $f$ of the right hand side of (\ref{a-phi}):
\begin{equation}
I = I_0 \, {\rm exp } \left[ - \int \frac{n \, d \varphi}{\sqrt{2 \epsilon \left( \varphi \right)} M_p} \right] \;\;\;\Rightarrow\;\;\;
\langle I \rangle \propto a^n
\label{I-V}
\end{equation}
where $I_0$ can be chosen so that $I=1$ after    inflation.  As a concrete example, the choice
\begin{equation}
V  =  \frac{1}{2} \, m^2 \varphi^2  \;\;,\;\; I = {\rm e}^{\frac{c \, \varphi^2}{2  M_p^2}}
\label{example-VI}
\end{equation}
results in $\langle I \rangle \propto a^{-2 c}$. We thus see that a constant $\vec{E}^{(0)} $ is achieved for $c=1$. Many other choices of $V$ and $I$ are clearly possible, provided that their functional forms are related to each other to produce through 
(\ref{a-phi}) the desired time dependence for $\langle I \rangle$~\cite{Martin:2007ue}.

\subsection{Production of vector fluctuations from the $I^2 F^2$ term}

It is well known that, for $I=1$, the vector field is conformally coupled to a FRW background, and so its fluctuations are not excited by the expansion of the universe. On the contrary, as we now review, the choice $\langle I \rangle \propto a^{-2}$ that allows for a constant $\vec{E}^{(0)}$ solution, also excites the vector fluctuations to produce a classical and scale invariant spectrum of ``electric'' fluctuations at large scales. 

We  quantize the vector field in the Coulomb gauge $A_0 = 0$
\begin{eqnarray}
\vec{A} &=& \vec{A}^{(0)} + \sum_{\lambda = \pm } \int \frac{d^3 k}{\left( 2 \pi \right)^{3/2}} 
\, {\rm e}^{i \vec k \vec x} \, \vec{\epsilon}_\lambda \left( \vec{k} \right)  \frac{{\hat V}_\lambda}{\langle I \rangle} \nonumber\\ 
{\hat V}    & \equiv &   \, a_\lambda \left( \vec{k} \right) \, V_\lambda \left( k \right) + 
 a_\lambda^\dagger \left( - \vec{k} \right) \, V_\lambda^* \left( k \right)    
\label{gauge-quant}
\end{eqnarray}
where $\vec{\epsilon}_\lambda$ are circular polarization vectors satisfying the relations $\vec{k}\cdot \vec{\epsilon}_{\pm} \left( \vec{k} \right) = 0$, $\vec{k} \times \vec{\epsilon}_{\pm} \left( \vec{k} \right) = \mp i k \vec{\epsilon}_{\pm} \left( \vec{k} \right)$,
$\vec{\epsilon}_\pm \left( \vec{-k} \right) = \vec{\epsilon}_\pm \left( \vec{k} \right)^*$, and normalized according to $\vec{\epsilon}_\lambda \left( \vec{k} \right)^* \cdot \vec{\epsilon}_{\lambda'} \left( \vec{k} \right) = \delta_{\lambda \lambda'}$. The annihilation / creation operators satisfy $\left[ a_\lambda \left( \vec{k} \right) ,\, a_{\lambda'}^\dagger \left( \vec{k}' \right) \right] = \delta_{\lambda \lambda'} \, \delta^{(3)} \left( \vec{k} - \vec{k}' \right)$.

The mode functions satisfy the evolution equation
\begin{equation}
V_\lambda'' + \left( k^2 - \frac{\langle I \rangle''}{\langle I \rangle} \right) V_\lambda = 0
\end{equation}
where prime denotes derivative with respect to conformal time $\tau$. For $\langle I \rangle \propto a^{-2} \propto \tau^2$ (we disregard slow roll corrections, so that $a = - \frac{1}{H \tau}$),  the properly normalized vector modes are
\begin{eqnarray}
V_\lambda \simeq \frac{1 + i  k \tau}{\sqrt{2} k^{3/2} \,    \tau   } \, {\rm e}^{-i k \tau}
\label{V-sol}
\end{eqnarray}

We Fourier transform the ``electric'' and ``magnetic'' fields (\ref{EB-ratra})
\begin{eqnarray}
\vec{E} & = & \vec{E}^{(0)} +  \int \frac{d^3 k}{\left( 2 \pi \right)^{3/2}} 
\, {\rm e}^{i \vec k \vec x} \, \delta \vec{E} \left( {\vec k} \right)
\nonumber\\
\vec{B} & = &   \int \frac{d^3 k}{\left( 2 \pi \right)^{3/2}} 
\, {\rm e}^{i \vec k \vec x} \, \delta \vec{B} \left( {\vec k} \right)
\end{eqnarray}

Inserting the solutions  (\ref{V-sol}) in  (\ref{EB-ratra}) we see that the   ``electric'' and ``magnetic''  fields  become classical (commuting) fields at super-horizon scales
\begin{eqnarray}
 \delta \vec{E} \left( {\vec k} \right) &=& 
\sum_\lambda {\cal E}_k  \;  \vec{\epsilon}_\lambda  \left( \vec{k} \right)  \left[     a_\lambda \left( \vec{k} \right)   +   a_\lambda^\dagger  \left( - \vec{k} \right)  \right] \nonumber\\
 \delta \vec{B} \left( {\vec k} \right) &=& \sum_\lambda {\cal B}_k  \; \lambda \,  \vec{\epsilon}_\lambda  \left( \vec{k} \right)  \left[     a_\lambda \left( \vec{k} \right)   +   a_\lambda^\dagger  \left( - \vec{k} \right)  \right] \nonumber\\
 {\cal E}_k  &  \simeq &  \frac{3 H^2}{\sqrt{2} k^{3/2}} \;\;,\;\;
 {\cal B}_k  \simeq  \frac{ H^2 \, \tau}{\sqrt{2} k^{1/2}} \;\;,\;\; - k \tau \ll 1
\label{EB-sol}
\end{eqnarray}
Namely, the ``electric'' field fluctuations are nearly constant outside the horizon, while the ``magnetic'' field fluctuations rapidly decrease. As we mentioned, the electric field fluctuations are scale invariant (slow roll corrections will slightly tilt their spectrum; however, we disregard slow roll corrections in this work whenever compared with a non-vanishing expression at $0-$th order in slow roll).

Finally, we note that this mechanism enjoys a duality symmetry  $\langle I \rangle \leftrightarrow \frac{ 1}{ \langle I \rangle }$. Under this exchange, the ``electric'' and ``magnetic'' modes interchange their role, $\left\vert \delta \vec{E} \right\vert^2 \leftrightarrow \left\vert \delta \vec{B} \right\vert^2 $. The original mechanism  \cite{Ratra:1991bn}  aims to produce  fluctuations with scale invariant magnetic energy, and therefore has  $\langle I \rangle \propto a^2$ during inflation. This corresponds to choosing $c=-1$ in the example (\ref{example-VI}). For definiteness, our explicit computations are done for   $\langle I \rangle \propto a^{-2}$. However, our results can be readily extended to  the context of  \cite{Ratra:1991bn} by exploiting this duality.

\subsection{Classical anisotropy  when the CMB modes leave the horizon} \label{sec:IR}

We denote by $\vec{E}^{(0)}$ the ``electric'' field  obtained from solving the classical equations of motion of a given model. For instance, as we discuss in Subsection \ref{subsec:Watanabe},  the classical equations of motion of the anisotropic inflationary model \cite{Watanabe:2009ct}, characterized by $c \simeq 1$ in  (\ref{example-VI}),  admit an attractor solution with a nearly constant  $\vec{E}^{(0)}$.

We denote by $\vec{E}_{\rm classical}$ the classical and  homogeneous electric field measured by a local observer at some time $\tau$ during inflation. We can assume that at the initial time of inflation  $\tau_{\rm in}$ no classical fluctuations are present, so that  $\vec{E}_{\rm classical} =   \vec{E}^{(0)}$ at $\tau_{\rm in}$. However, this identification is no longer exactly true at any later time. Indeed, at the time $\tau > \tau_{\rm in}$ during inflation
\begin{equation}
\vec{E}_{\rm classical} =   \vec{E}^{(0)} +  \vec{E}^{\rm IR} 
\label{E-classical}
\end{equation}
where the second quantity (IR = ``infra-red'') denotes the sum  of all the  modes $\delta \vec{E}_k$ that left the horizon between the times $\tau_{\rm in}$ and $\tau$. These modes have become classical and are homogeneous from the point of view of a local observer present at $\tau$.

The same considerations apply to the ``magnetic''  component of the vector field. For any single realization of the mechanism
(\ref{L-ratra}) with $\langle I \rangle \propto a^{-2}$,  the quantities $  \vec{E}^{\rm IR} $ and $  \vec{B}^{\rm IR} $ are drawn by, respectively, a gaussian  (to very good approximation) statistics, with vanishing mean and with variance
\begin{eqnarray}
\sigma^2_{ \vec{E}^{\rm IR},N}    & = & \langle \delta \vec{E} \left( \vec{x} \right)^2 \rangle = \frac{1}{ \pi^2} \int_{\rm IR} d k \vert {\cal E}_k \vert^2 \simeq \frac{9 H^4}{2 \pi^2} \int_{\rm IR} \frac{d k}{k} \nonumber\\ 
& = &  \frac{9 H^4}{2 \pi^2} \, N \nonumber\\
\sigma^2_{ \vec{B}^{\rm IR}}  & = &  \dots =  \frac{3 H^4}{8 \pi^2}
\label{variance-EIR}
\end{eqnarray}
where the IR modes are those characterized by momentum $k$ in the interval $\frac{1}{-\tau_{\rm in}} < k < \frac{1}{-\tau}$, and where  $N$ is the number of e-folds of inflation from $\tau_{\rm in}$ and $\tau$.

These quantities  are the natural expectation for the energy that gets progressively stored in the ``electromagnetic'' field during inflation
\begin{equation}
\rho_{\delta E,N} = \frac{ \langle \delta \vec{E} \left( \vec{x} \right)^2 \rangle }{ 2 } \simeq   \frac{9 H^4}{4 \pi^2} \, N 
\;\;\;,\;\;\; \rho_B \ll \rho_{\delta E} 
\label{rho-dE}
\end{equation}
This has been well appreciated in the magnetogenesis applications of this mechanism (we remark that the role of the ``electric'' and ``magnetic'' energy is interchanged for $\langle I \rangle \leftrightarrow \frac{1}{\langle I \rangle}$). For example, ref. \cite{Demozzi:2009fu} computed the energy density accumulated in these super-horizon modes for all possible values of $n$ in the $\langle I \rangle \propto a^{n}$ dependence. For $\vert n \vert > 2$, this energy density grows as $a^{4\left( \vert n / 2 \vert - 1 \right)}$  (while $n=\pm2$ results in   $\rho \propto \ln a = N$, as we have seen), until the backreaction of this energy is no longer negligible. Ref. \cite{Kanno:2009ei} studied the regime of strong backreaction that takes place for $n>2$. Ref.  \cite{Fujita:2012rb} studied  magnetogenesis for a more general time dependence of $I$, imposing as one of the conditions that the energy density in the classical super-horizon modes remains less than that of the inflaton. 

We obtain a first  limit on the total duration of inflation $N_{\rm tot}$  by imposing that  $  \vec{E}^{\rm IR} $ has a negligible effect on the background evolution. The strongest backreaction constraint  does not actually come from $\rho_{\delta E} \ll V$, but rather from the evolution equation of the inflaton (loosely speaking, it is ``easier'' for the vector field to affect the motion of the inflaton, that is slowly rolling on a flat potential, than the expansion rate). The corresponding condition is   $\rho_{\delta E} \ll 2 \, \epsilon \, H^2 \, M_p^2$, and it is satisfied for  $N_{\rm tot}  \ll {\rm O } \left( 10^7 \right)$ \cite{Barnaby:2012tk}.

The variances of the three components of $\vec{E}^{\rm IR}$ are equal to each other,~\footnote{
Actually, one can obtain a small difference proportional to the small difference $\Delta H$ in the expansion rates of the different directions. As we well see, $\Delta H / H $ needs to be $\la 10^{-8}$, and therefore this difference is completely negligible for the present discussion.} and equal to $1/3$ of the value given in (\ref{variance-EIR}).  This, however, does not mean that  in a given realization the super-horizon modes  add up to equal amounts in all three directions. Indeed, the difference between different directions is drawn from the statistics 
\begin{eqnarray}
& &
\left\langle \delta E_x^2 -  \delta E_y^2 \right\rangle = 0 \nonumber\\
& &
\sqrt{\left\langle \left(   \delta E_x^2 -  \delta E_y^2      \right)^2 \right\rangle} = 2 \left\langle \delta E_x^2 \right\rangle =
\frac{4}{3} \, \rho_{\delta E,N}
\end{eqnarray}
Therefore, $\rho_{\delta E}$ is also the typical amount of the classical anisotropy provided by the IR modes. This is not surprising since, in any given realization, $\vec{E}^{\rm IR}$ is a classical vector that points in some given direction.

Only for $\vert  \vec{E}^{(0)} \vert \gg \vert   \vec{E}^{\rm IR} \vert$, or, equivalently, for $\rho_{E^{(0)}} \gg \rho_{\delta_E}$, 
the classical electric field measured by the observer is (deterministically) given by the solution of the classical equations of motion.  If this is not the case, one should conclude that  the solution of the classical equations of motion is unstable under quantum corrections, and one should expect large corrections to the predictions made if only  $\vert  \vec{E}^{(0)} \vert$ is considered.\footnote{Before eq. (\ref{E-classical}), we set $\vec{E}_{\rm classical} =   \vec{E}^{(0)}$ at the start of inflation. We note that if this is not the case the departure of the classical electric field from the solution of the classical equation of motion will  in general be even greater.} In our computations below, both the contributions to $\vec{E}_{\rm classical} $ will be accounted for (as we will see, this amounts in considering loop contributions to the power spectrum and the anisotropic spectrum of the cosmological perturbations).

\section{Anisotropic source of the Cosmological Perturbations}
\label{sec:perturbations}

As we have discussed in length in the previous Section, for the mechanism we are studing  the cosmological perturbations that we observed experienced a classical homogeneous vector field   (\ref{E-classical})  when they left the horizon. This breaks the background isotropy, and, strictly speaking, the local patch where these modes live has a Bianchi-I geometry with residual $2$d isotropy
\begin{equation}
d s^2 = - d t^2 + a^2 \left( t \right) d x^2 +  b^2 \left( t \right) \left[  d y^2 + d z^2 \right]
\label{bianchi}
\end{equation}
where for the definiteness the $x-$axis has been oriented along $\vec{E}_{\rm classical}$.

To characterize the anisotropy,  we define
\begin{equation}
\frac{\Delta H}{H} \equiv \frac{3 \left( H_y - H_x \right)}{H_x + H_y + H_z} 
\label{dH-H-def}
\end{equation}
This nearly corresponds to the mechanism of anisotropic inflation of   \cite{Watanabe:2009ct}, where the model (\ref{example-VI}) has been employed, and where it is shown that the classical equations of the model admit an anisotropic solution with $\frac{\Delta H}{H} \simeq \left( c - 1 \right) \epsilon$. In \cite{Watanabe:2009ct}, and in the successive works \cite{Himmetoglu:2009mk,Dulaney:2010sq,Gumrukcuoglu:2010yc,Watanabe:2010fh} that study the linearized perturbations of this model, it is assumed that $\vec{E}_{\rm classical} = \vec{E}^{(0)}$, while, as we have discussed, these two quantities are in general different. Therefore, this model leads to anisotropic inflation even for $c=1$. Given this strong correspondence, our study has several relations with  \cite{Watanabe:2009ct}, and for example we will show that our results for the power spectrum coincides with that of \cite{Dulaney:2010sq,Gumrukcuoglu:2010yc,Watanabe:2010fh}, once the value of $ \vec{E}^{(0)}$ used there is replaced by the full  $\vec{E}_{\rm classical} $ (for the bispectrum instead, our result is completely new, since the perturbations of 
 \cite{Watanabe:2009ct} have been so far studied only at the linearized level).
 
 In the studies \cite{Himmetoglu:2009mk,Dulaney:2010sq,Gumrukcuoglu:2010yc,Watanabe:2010fh}, the perturbations are separated according to how  they transform under an SO(2) rotation in the symmetry plane orthogonal to the vector vev, as originally done in \cite{Gumrukcuoglu:2006xj,Gumrukcuoglu:2007bx}. This simplifies the problem, as modes that transform differently are not coupled to each other at the linearized level. Still, the non-FRW background results in a very involved computation, once the perturbations of all the fields (metric included) are taken into account. In particular, the anisotropy does not increase the number of physical modes, but results in couplings between these modes that would be absent in the FRW case \cite{Gumrukcuoglu:2006xj,Gumrukcuoglu:2007bx}. The main result of \cite{Dulaney:2010sq,Gumrukcuoglu:2010yc,Watanabe:2010fh}, which is somewhat surprising a-priori, is that an anisotropic parameter $g_* = {\rm O } \left( 0.1 \right)$ is obtained for a $\frac{\Delta H}{H} = {\rm O } \left( 10^{-8} \right)$ background anisotropy. Motivated by this result, one can instead use a different approach, and perform  the standard quantization of the cosmological perturbations on a FRW background, ignoring at zeroth order the couplings between the different modes. Such couplings can be taken into account as perturbative mass insertions in the in-in formalism. This is effectively the procedure adopted in  \cite{Dulaney:2010sq,Watanabe:2010fh} when they  solve analytically the linearized theory. Moreover ref.  \cite{Watanabe:2010fh} showed that the dominant operator that determines $g_*$ in this perturbative evaluation is the $\delta \varphi - \delta A_\mu$ coupling obtained from expanding the vector kinetic term, and with no contribution from the metric perturbations. This analytic approximated result is in excellent agreement  \cite{Watanabe:2010fh}   with the one obtained from an exact numerical evolution of $g_*$  (in which the full quadratic action of all the $2$d scalar modes is retained). Moreover, the analytical and numerical results of   \cite{Watanabe:2010fh} agree with those of  \cite{Dulaney:2010sq} and \cite{Gumrukcuoglu:2010yc}, respectively. 

We use the same computational scheme for  the bispectrum computation. Specifically, we disregard metric perturbations, and we use in the in-in formalism the $0th$ order eigenmodes obtained from the approximate FRW quantization.  The bispectrum is produced by interactions, and we know that for slow roll FRW inflation the interactions of the metric perturbations produce an 
unobservable bispectrum. We will see instead that the  interaction between the vector and the scalar field, which is encoded in the vector kinetic term, results in a larger, and potentially observable, signal. Ref. \cite{Barnaby:2012tk} proved this explicitly  for the case of $\vec{E}_{\rm classical} = 0$, by solving the second order equation for  the curvature perturbation  in spatially flat gauge $\zeta= - \frac{\cal H}{\varphi'} \, \delta \varphi$ (we recall that prime denotes derivative with respect to conformal time $\tau$, while ${\cal H} = \frac{a'}{a}$). It was shown in  \cite{Barnaby:2012tk} that the contribution from the direct vector-scalar interaction  is slow-roll enhanced with respect to that coming from the interactions of the metric. There is no reason to expect that the relative strength of the effects should change for a  ${\rm O } \left( 10^{-8} \right)$ background anisotropy.~\footnote{We note that metric perturbations are also disregarded in the computations  \cite{Yokoyama:2008xw,Karciauskas:2008bc,Karciauskas:2011fp,Emami:2011yi,Lyth:2012br,Lyth:2012vn} of the anisotropic bispectrum through the waterfall mechanism.}

Therefore we retain the FRW quantization of the vector field performed in the previous Section. The remaining perturbations (of the scalar field and of the metric) are also quantized as in the standard FRW case. We then expand the vector kinetic term into $\delta A$ and $\delta \varphi$. We use the resulting interactions to determine  the dominant anisotropic contribution to the power spectrum of $\zeta$ and  the dominant  contribution to the bispectrum. 

Therefore, all the dominant effects arise from expanding the only interaction term between the inflaton and the vector field,
\begin{eqnarray}
\Delta {\cal L}  =  \frac{-a^4}{4} \!\!  \left( \langle  \frac{\delta I^2}{\delta \varphi } \rangle \delta \varphi + \frac{1}{2} \langle \frac{\delta^2 I^2}{\delta \varphi^2 } \rangle \delta \varphi^2  + \dots \right) \! \left(  \langle F_{\mu \nu} \rangle + \delta F^{\mu \nu} \right)^2 \nonumber\\
\label{L-int-exp}
 \end{eqnarray}

In spatially flat gauge,  $\delta \varphi = - \frac{\varphi'}{\cal H} \, \zeta = \sqrt{2 \epsilon} \, M_p \, \zeta$. We note that, in principle, $\zeta$ has additional contributions proportional to the perturbations of the vector field. Ref.  \cite{Barnaby:2012tk}  showed that these contributions are completely subdominant in the case of $\vec{E}_{\rm classical} = 0$.  We believe that it is very natural to assume that this continues to be the case also in the current context. Indeed, as we shall see, $\frac{\rho_E}{\rho_\varphi}  $ needs to be $\la {\rm O } \left( 10^{-8} \right)$ during inflation or otherwise the power spectrum is too anisotropic.
 This ratio further decreases between the end of inflation and the inflaton decay, when the inflation field performs coherent oscillations, so that $\rho_\varphi \propto a^{-3}$, while the vector kinetic term becomes standard, and $\rho_E \propto a^{-4}$. If this is the case,
\begin{equation}
\frac{a_{\rm end \, infl}}{a_{\rm reh}} \simeq 10^{-10} \, \left( \frac{T_{\rm reh}}{10^9 \, {\rm GeV}} \right)^{4/3} \, \left( \frac{10^{15} \, {\rm GeV} }{ H_{\rm end \, infl} } \right)^{2/3}
\end{equation}
where $T_{\rm reh}$ and $a_{\rm reh}$ are, respectively, the temperature of the inflation decay products and the scale factor at the inflaton decay. We see that it is therefore natural to disregard $\rho_E$ and $\delta \rho_E$ at reheating. In this way, the only relevant contribution of $\delta A$ to the final curvature perturbation is the modification of $\delta \varphi$ induced by the vector-inflaton coupling (which is precisely the effect that we are computing). We note that this   assumption is also made in 
\cite{Dulaney:2010sq,Gumrukcuoglu:2010yc,Watanabe:2010fh} when they give the power spectrum of $\zeta$ in the model  \cite{Watanabe:2009ct}.

Using the expression (\ref{I-V}), we have, for the first two terms in the expansion of $I^2$,
\begin{eqnarray}
\left\langle \frac{\partial I^2}{\partial \varphi} \right\rangle \delta \varphi  & = & 
- 2 n \, \left\langle \frac{I^2}{\sqrt{2 \epsilon} M_p} \right\rangle \delta \varphi = 
 - 2 n \left\langle I^2 \right\rangle \zeta
\nonumber\\
\frac{1}{2} \, \left\langle \frac{\partial^2 I^2}{\partial \varphi^2} \right\rangle \delta \varphi ^2 & = & 
\left\langle \left[ \frac{ n^2}{\epsilon M_p^2} - \frac{n}{M_p^2} \left( 1 - \frac{\eta}{2 \epsilon} \right) \right] I^2 \right\rangle \delta \varphi^2 \nonumber\\
& &  \simeq
 2 n^2  \left\langle I^2 \right\rangle \zeta^2
\end{eqnarray}
where $\eta = M_p^2 \frac{V''}{V}$ is a slow roll parameter, and where in the final approximation we retained only the dominant term in slow roll approximation. 

We inserted these expressions in   (\ref{L-int-exp}), taking $n=-2$ (which corresponds to a constant ``electric'' field).  We expanded also the vector part as in (\ref{gauge-quant}), and we  computed the  contributions to the power spectrum and the bispectrum  combining the resulting vertices. We verified that the leading diagrams do not contain interactions coming from  the second order term $ \frac{\delta^2 I^2}{\delta \varphi^2 }  $  (if this was not the case, we should worry about the convergence of the $I \left[ \langle \varphi \rangle + \delta \varphi \right]$ expansion). Therefore we have the two dominant  $\propto \delta \varphi  A^{(0)} \, \delta A$ and $\propto \delta \varphi \delta A^2$ interactions~\footnote{When we perform the expansion, linear terms in the perturbations are removed by the background equations of motion; we commit a mistake by disregarding the effect of the gauge field vev in the background evolution, but, as we remarked, we work in a regime where this effect is negligible.} 
\begin{eqnarray}
 {\cal L}_{\rm int}  
& \supset &  a^4 \left[  4 E_x^{(0)} \delta E_x \zeta   + 2  \, \delta \vec{  E } \cdot  \delta \vec{  E } \,  \zeta  \right] \nonumber\\
 &  \equiv &  {\cal L}_{\rm int,1} + {\cal L}_{\rm int,2}  
\label{Lint}
\end{eqnarray}

We note that  the vector field enters in the quadratic term ${\cal L}_{\rm int,1}$   only in the combination $ \propto 2 \langle E_x \rangle \, \delta E_x \subset \vec{E}^2 - \vec{B}^2 \propto F_{\mu \nu} F^{\mu \nu}$. We also note that the cubic term 
 ${\cal L}_{\rm int,2}$  should also have the term $- 2 a^4 \delta \vec{B} \cdot \delta \vec{B} \, \zeta$; this term however 
gives a subdominant contribution to $\zeta$ with respect to contribution from the ``electric'' components, since the ``electric'' modes are much greater than the ``magnetic'' ones at super-horizon scales - see eq. (\ref{EB-sol}) - which is where they provide the greatest contribution to $\zeta$ (as discussed in \cite{Barnaby:2012tk} and below). 

We stress that, although in this Section we have often mentioned the anisotropic inflationary  model  \cite{Watanabe:2009ct}, this interaction lagrangian applies to all models that verify the two assumptions spelled out at the beginning of the introduction. We actually see that the precise functional forms of $I$ and $V$ do not enter in (\ref{Lint}), apart from the fact that the relation (\ref{I-V}) has been imposed, with $n=-2$. Identical results would be obtained for $n=2$, exploiting the ``electric'' -- ``magnetic'' duality of the mechanism. We recall that $n=\pm2$ precisely  correspond to enforcing the assumption 2. made at the beginning of the Introduction. For the choice (\ref{example-VI}), one has $n=-2c$. However, we stress that (\ref{Lint}) is valid independently of this choice.

The interaction lagrangian  (\ref{Lint})  enters in the $n-$point correlation functions through the in-in formalism relation
\begin{eqnarray}
&& \left\langle {\hat \zeta}_{\vec{k}_1} \,  {\hat \zeta}_{\vec{k}_2} \dots {\hat  \zeta}_{\vec{k}_n} \left( \tau \right)  \right\rangle = \sum_{N=0}^\infty \left( - i \right)^N \int^\tau  d \tau_1 \dots   \int^{\tau_{N-1}}  d \tau_N \nonumber\\
&& \;\;\;  \left\langle \left[ \left[ \dots  \left[ {\hat \zeta}_{\vec{k}_1}^{(0)} \,  {\hat \zeta}_{\vec{k}_2}^{(0)} \dots  {\hat \zeta}_{\vec{k}_n}^{(0)} \left( \tau \right) ,\;
H_{\rm int} \left( \tau_1 \right) \right] , \dots \right] ,\, H_{\rm int} \left( \tau_N \right) \right] \right\rangle \nonumber\\
\label{in-in}
 \end{eqnarray} 
 where in our computational schemes the quantity $ {\hat \zeta}_{\vec{k}}^{(0)} $ is the Fourier transform of the (unperturbed) FRW quantized field
\begin{eqnarray}
\zeta^{(0)} = \int \frac{d^3 k}{\left( 2 \pi \right)^{3/2}} {\rm e}^{i \vec{k} \cdot \vec{x}} \, {\hat \zeta}_{\vec{k}}^{(0)}
\;\;\;,\;\;\;  {\hat \zeta}_{\vec{k}}^{(0)} \equiv 
 \zeta_{\vec{k}}^{(0)} \,  a_{\vec{k}}  +  \zeta_{\vec{k}}^{(0) \, *} \,  a^\dagger _{-\vec{k}} \nonumber\\
\label{zeta-0}
 \end{eqnarray}
At leading order in slow roll, we have
\begin{eqnarray}
\zeta_k^{(0)} \left( \tau \right) \simeq \frac{H \left( 1 + i \, k \, \tau \right)}{2 \sqrt{\epsilon} M_p \, k^{3/2}} \, {\rm e}^{-i k \tau }
 \label{zeta-0-modes}
\end{eqnarray}

The unperturbed fields are also employed in the interaction hamiltonian $H_{\rm int} \left( \tau \right) = - \int d^3 x \, {\cal L}_{\rm int} \left( \tau ,\, \vec{x} \right)$. The two terms in (\ref{Lint}) give, respectively, rise to the two terms
\begin{eqnarray}
H_{{\rm int},1} \left( \tau \right) & = & - \frac{4  \,  E_x^{(0)}   }{H^4 \tau^4} \int d^3 k \, \delta E_x \left( \tau ,\,  \vec{k} \right) \, {\hat \zeta}_{-\vec{k}}^{(0)} \left( \tau \right) \nonumber\\
H_{{\rm int},2} \left( \tau \right) & = & - \frac{2}{H^4 \tau^4}  \!\! \int \!\! \frac{  d^3 k  d^3 p }{ \left( 2 \pi \right)^{3/2} }  \delta \vec{E}  \left( \tau ,\,  \vec{k} \right) 
\!\cdot\! \delta \vec{E}  \left( \tau ,\,  \vec{p} \right) \, {\hat \zeta}_{-\vec{k}-\vec{p}}^{(0)} \left( \tau \right) \nonumber\\
\label{Hint}
\end{eqnarray}

Once inserted into (\ref{in-in}), the two interaction terms  give rise to the leading contributions to the power spectrum and bispectrum  described by the diagrams shown, respectively,   in Figures (\ref{fig:zz}) and (\ref{fig:zzz}). These quantities are computed in the following Section.

\section{Anisotropic power spectrum}
 \label{section:zz}

\begin{widetext}
\begin{figure*}[ht!]
\centerline{
\includegraphics[width=\textwidth,angle=0]{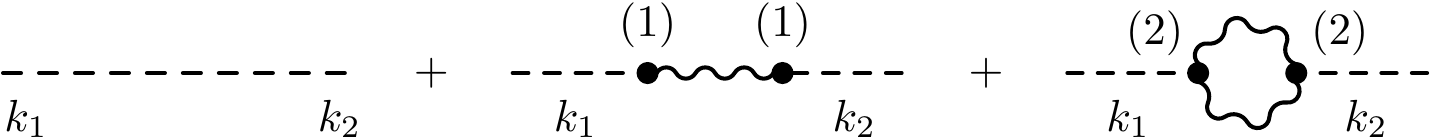}
}
\caption{Leading diagrams for $\langle \zeta^2 \rangle$, with the vertices labelled as in (\ref{Hint}).
}
\label{fig:zz}\end{figure*}
\end{widetext}

The total power spectrum $P_\zeta \left( \vec{k} \right)$ is related to the full two-point correlation function (\ref{in-in}) by
\begin{equation}
\left\langle {\hat \zeta}_{\vec{k}_1} \,  {\hat \zeta}_{\vec{k}_2} \right\rangle = 2 \pi^2  \, \frac{\delta^{(3)} \left( \vec{k}_1 + \vec{k}_2 \right)}{k_1^3} \,  P_\zeta \left( \vec{k}_1 \right) 
\end{equation}

We denote the  contributions of the first and of the last two diagrams  in  Figure \ref{fig:zz} as, respectively,
\begin{equation}
\left\langle  {\hat \zeta}_{\vec{k}_1} \,  {\hat \zeta}_{\vec{k}_2} \right\rangle = 
\left\langle  {\hat \zeta}_{\vec{k}_1}^{(0)} \,  {\hat \zeta}_{\vec{k}_2}^{(0)} \right\rangle + 
\delta \left\langle  {\hat \zeta}_{\vec{k}_1} \,  {\hat \zeta}_{\vec{k}_2}\right\rangle  
\end{equation}
and, correspondingly,
\begin{equation}
  P_\zeta = {\cal P}^{(0)} + \delta P
\end{equation}

These quantities are computed in the following  Subsections.

\subsection{Tree level contributions} \label{subsection:zz-tree}

The first diagram in Figure \ref{fig:zz} gives the (unperturbed) FRW power spectrum
\begin{equation}
{\cal P}^{(0)} = \frac{H^2}{8 \pi^2 \epsilon M_p^2} = \frac{H^4}{4 \pi^2 \dot{\varphi}^2}
\end{equation}
at super-horizon scales, where the slow roll expression (\ref{zeta-0-modes}) has been used.

The second diagram in  Figure \ref{fig:zz} gives the anisotropic contribution
\begin{eqnarray}
\delta \langle {\hat  \zeta}_{{\vec k}_1} {\hat   \zeta}_{{\vec k}_2  } \left( \tau \right)  \rangle \vert_{1}  
& = & - \,  \int_{\tau_{\rm min}}^\tau d \tau_1  \int_{\tau_{\rm min}}^{\tau_1} d \tau_2 \nonumber\\
& & \!\!\!\!\!\!\!\!
\left \langle \left[ \left[ {\hat  \zeta}_{{\vec k}_1}^{(0)}  {\hat  \zeta}_{{\vec k}_2  }^{(0)} \left( \tau \right)  ,  {\cal H}_{{\rm int},1} \left( \tau_1 \right) \right] ,\,     {\cal H}_{{\rm int},1} \left( \tau_2 \right) \right] \right \rangle \nonumber\\
\label{inin-zz-2}
\end{eqnarray}
We are interested in the power spectrum at super horizon scales, $ k \vert \tau \vert \ll 1$.  The time integral (\ref{inin-zz-2}) is  dominated by the times for which the modes in $H_{\rm int} \left( \tau_i \right)$ are also outside the horizon (mathematically, the contribution in the sub-horizon phase is suppressed by the oscillatory phases in the mode functions). This condition will be relevant for setting  $\tau_{\rm min}$ (see below). As remarked before eq. (\ref{EB-sol}), in this regime the vector field is classical, and does not contribute to the commutators. The only nontrivial elements in the commutators are therefore
\begin{eqnarray}
& & \left[ {\hat \zeta}_{\vec k}^{(0)} \left( \tau \right) ,\,  {\hat \zeta}_{{\vec k}'}^{(0)} \left( \tau' \right) \right] \nonumber\\
& & \quad\quad\quad =   
\left( \zeta_k^{(0)} \left( \tau \right)  \zeta_k^{(0)*} \left( \tau' \right) - {\rm c. c.} \right) \, \delta^{(3)} \left( \vec{k} + \vec{k}' \right) \nonumber\\
& & \quad\quad\quad      \simeq \frac{- i H^2 \left[  \tau^3  -   \tau^{'3} \right] }{6 \epsilon M_p^2}
\, \delta^{(3)} \left( \vec{k} + \vec{k}' \right)
\end{eqnarray}
where the last result is true in the super-horizon regime. We insert (\ref{Hint}) into (\ref{inin-zz-2}) and perform the commutators between the ${\hat \zeta}^{(0)}$ fields. The two resulting $\delta-$functions are employed to perform the two integrals over momenta, leading to
\begin{eqnarray}
\delta \langle {\hat  \zeta}_{{\vec k}_1} {\hat   \zeta}_{{\vec k}_2  } \left( \tau \right)  \rangle  \vert_{1}  
& \simeq &  \frac{4 \, E_x^{(0) 2}}{9 \epsilon^2 M_p^4 H^4} 
 \prod_{i=1}^2  \int_{\tau_{\rm min}}^\tau \frac{d \tau_i}{\tau_i^4} \left[   \tau^3 -   \tau_i^3 \right] \nonumber\\
& & \quad \left\langle \delta E_x \left( \tau_1 ,\, \vec{k}_1 \right)  \delta E_x \left( \tau_2 ,\, \vec{k}_2 \right) \right\rangle
\label{partial-zz2}
\end{eqnarray}
Requiring that the vector field in this expression is in the super-horizon regime limits each time integral to $\tau_i > - \frac{1}{k_i}$. 
This sets the value of $\tau_{\rm min}$ in the two integrals. Using the expressions (\ref{EB-sol}), and the identity 
\begin{equation}
\sum_\lambda \epsilon_{\lambda,i} \left( \vec{k} \right) \,  \epsilon_{\lambda,j}^* \left( \vec{k} \right) = \delta_{ij} - {\hat k}_i \, {\hat k}_j
\end{equation}  
we obtain
\begin{eqnarray}
\delta \langle {\hat  \zeta}_{{\vec k}_1} {\hat   \zeta}_{{\vec k}_2  } \left( \tau \right)  \rangle  \vert_{1}  & \simeq & 
 \frac{2 E_x^{(0) 2}  }{\epsilon^2 \, M_p^4} \, \frac{\delta^{(3)} \left( \vec{k}_1 + \vec{k}_2 \right)}{k_1^3} \, \sin^2  \theta_{{\hat k}_1 , {\hat E }^{(0)}  } \nonumber\\
 & & \quad\quad \quad\quad
   \left\{  \int_{- \frac{1}{k_1} }^\tau \frac{d \tau'}{\tau'^4} \left[   \tau^3 -   \tau^{'3} \right] \right\}^2 
\end{eqnarray}

Changing variable $y' \equiv \frac{\tau'}{\tau}$, and recalling that $-k_1 \tau \ll 1$, the time integral in the second line becomes
\begin{equation}
\int_1^{-\frac{1}{k_1 \, \tau}} d y' \, \frac{y'^3-1}{y^{' 4}} \simeq \ln \frac{1}{-k_1 \, \tau}
\end{equation}
At the end of inflation this quantity becomes $N_{k_1}$, namely the number of e-folds before the end of inflation at which the modes with wavenumber $k_1$ left the horizon.  Using this result, 
\begin{equation}
 \delta P_{1} \left( \tau_{\rm end} ,\, \vec{k}  \right) \simeq \frac{24}{\epsilon} \, \frac{E_x^{(0) 2}}{V \left( \varphi \right) } \, N_k^2 {\cal P}^{(0)}   \sin^2  \theta_{{\hat k} , {\hat E }^{(0)}  }   
\label{dP1-aniso}
\end{equation}
where $\tau_{\rm end}$ denotes the end of inflation (we assume that $I$ rapidly approaches $1$ at the end of inflation, so that the vector field is rapidly diluted away by the expansion, and the power spectrum of $\zeta$ freezes out).

For the anisotropic inflationary  model of  \cite{Watanabe:2009ct},  eq. (\ref{dP1-aniso})  coincides with the analytic result of \cite{Dulaney:2010sq,Watanabe:2010fh}, which, as we remarked, is in excellent agreement with the full numerical computation of  \cite{Gumrukcuoglu:2010yc,Watanabe:2010fh}. This confirms the validity of all the approximations that we have performed.

\subsection{Including the Loop contribution} \label{subsection:zz-loop}

The expression for the  loop diagram in  Figure \ref{fig:zz} is analogous to eq. (\ref{inin-zz-2}), with $ {\cal H}_{{\rm int},2} $ replacing $  {\cal H}_{{\rm int},1} $. We perform the commutators, again keeping in mind that the time integrals are dominated by fields in the super-horizon regime. We obtain
\begin{eqnarray}
& & 
\delta \langle {\hat  \zeta}_{{\vec k}_1} {\hat   \zeta}_{{\vec k}_2  } \left( \tau \right)  \rangle  \vert_{2}   \simeq  
\frac{1}{9 \epsilon^2 H^4 M_p^4} \!
  \int \frac{d^3 p d^3 q}{\left( 2 \pi \right)^3}   \prod_{i=1}^2  \! \int_{\tau_{\rm min}}^\tau \!\!  \frac{d \tau_i}{\tau_i^4} \left[   \tau^3 -   \tau_i^3 \right]  \nonumber\\
& &  \left\langle \delta E_i \left( \tau_1 ,\, \vec{p} \right)  \delta E_i \left( \tau_1 ,\, \vec{k}_1 - \vec{p} \right)
  \delta E_j \left( \tau_2 ,\, \vec{q} \right)   \delta E_j \left( \tau_2 ,\, \vec{k}_2 - \vec{q} \right) \right\rangle \nonumber\\
\label{partial-zz3}
\end{eqnarray}

Evaluating the full correlator (disregarding the disconnected contraction) gives 
\begin{eqnarray}
& & \left( \delta \langle {\hat  \zeta}_{{\vec k}_1} {\hat   \zeta}_{{\vec k}_2  } \left( \tau \right)  \rangle  \vert_{2} \right)_{\rm theory}
  \simeq  
\frac{9 H^4 \delta^{(3)} \left( \vec{k}_1 + \vec{k}_2 \right)}{2 \epsilon^2 M_p^4} 
\nonumber\\
& & \quad\quad\quad\quad \int  \frac{d^3 p }{\left( 2 \pi \right)^3} \frac{1+\cos^2 \theta_{{\hat p},\widehat{k_1-p}}}{p^3 \, \vert \vec{k}_1 - \vec{p} \vert^3} 
\,  \prod_{i=1}^2  \int_{\tau_{\rm min}}^\tau \frac{d \tau_i}{\tau_i^4} \left[   \tau^3 -   \tau_i^3 \right]  \nonumber\\
\label{partial2-zz3}
\end{eqnarray}
where also in this case we recall that the main contribution comes from the times when all the fields are in the superhorizon regime (for the same reasons mentioned after eq. (\ref{inin-zz-2})). This corresponds to $\tau_{\rm min} = {\rm Max} \left[ - \frac{1}{p} ,\,- \frac{1}{\vert \vec{k}_1 - \vec{p} \vert} \right]$. Performing the time integrals, and keeping only the logarithmically enhanced term, leads to
\begin{eqnarray}
& & \left(  \delta \langle {\hat  \zeta}_{{\vec k}_1} {\hat   \zeta}_{{\vec k}_2  } \left( \tau \right)  \rangle  \vert_{2}  \right)_{\rm theory}  \simeq   
\frac{9 H^4 \delta^{(3)} \left( \vec{k}_1 + \vec{k}_2 \right)}{2 \epsilon^2 M_p^4} \nonumber\\
& & \quad   \int  \frac{d^3 p }{\left( 2 \pi \right)^3} \frac{1+\cos^2 \theta_{{\hat p},\widehat{k_1-p}}}{p^3 \, \vert \vec{k}_1 - \vec{p} \vert^3} \, {\rm ln }^2 \left( {\rm Min } \left[ \frac{1}{- \tau \, p} ,\, \frac{1}{- \tau \, \vert \vec{k}_1 - \vec{p} \vert} \right] \right) \nonumber\\
\label{partial3-zz3}
\end{eqnarray}

The final momentum integral naively diverges logarithmically at the two poles  $\vec{p} \rightarrow \vec{0} ,\, \vec{k}_1$. We need however to impose that the   the fields $\delta \vec{E}$ in (\ref{partial-zz3}) were inside the horizon at the start of inflation (or they would not be excited by the mechanism described in the previous Section). Therefore $p ,\, \vert \vec{k}_1 - \vec{p} \vert > - \frac{1}{\tau_{\rm in}}$, where $\tau_{\rm in}$ denotes the conformal time at the  start of inflation. With this cut-off
\begin{equation}
 \delta P_{2} \left( \tau_{\rm end} ,\, \vec{k}  \right) \vert_{\rm theory}  \simeq  192 \, {\cal P}^{(0)2} N_k^2 \, \left( N_{\rm tot} - N_k  \right)
  \label{dP2-theory}
  \end{equation}
This result accounts only for the contribution of the two poles and therefore is accurate only as long as the  logarithmic enhancement encoded in the last term is $\gg1$. For $ N_{\rm tot} \simeq  N_k $ this last factor should be replaced by a ${\rm O } \left( 1 \right)$ factor.  The result (\ref{dP2-theory}) was first derived in  \cite{Barnaby:2012tk} using the Green function method.

We note that the modes contributing to the logarithmic enhancement are precisely the modes that left the horizon during the first $N_{\rm tot} - N_{\rm CMB}$ e-folds of inflation, and that contribute to the classical field  $\vec{E}^{\rm IR}$. As we discussed 
after eq. (\ref{E-classical}), this quantity adds up with the solution $\vec{E}^{(0)}$ of the classical equations of motion to give
  $\vec{E}_{\rm classical}$, which is the homogeneous and classical ``electric'' field observed by the CMB modes. The quantity
(\ref{dP2-theory}) is the theoretical expectation value associated with the loop diagram; it is the result that one would obtain if one could average over several realizations of the mechanism (see also~\cite{Bartolo:2007ti}). However (assuming that this mechanism describes our universe)  the CMB modes exit the horizon after a single realization of the first $N_{\rm tot} - N_{\rm CMB}$ e-folds of inflation. They therefore are not affected by the theoretical average   $\langle \vert \vec{E}^{\rm IR} \vert\;^2 \rangle$ (which is statistically isotropic), but by the value that  $\vec{E}^{\rm IR}$ happened to assume in that single realization.

Therefore, if we want to compute the contribution of $\delta P_2$ to the  power spectrum that is observed  in a single realization, we need to replace the quantum operator of the IR modes entering in (\ref{partial-zz3}) with the classical Fourier transform of  $\vec{E}^{\rm IR}$. If we do so,  the expression (\ref{partial-zz3}) becomes formally identical to (\ref{partial-zz2}), with the quantity $ E_x^{(0) 2} \equiv \vert \vec{E}^{(0)} \vert^2 $ replaced by $\vert \vec{E}^{\rm IR} \vert^2$ and with $\delta E_x \left( \tau_i ,\, \vec{k}_i \right)$ (namely, the component of the fluctuations along the direction of $\vec{E}^{(0)}$) replaced by the component of  $\delta \vec{E}  \left( \tau_i ,\, \vec{k}_i \right)$ in the direction of  $\vert \vec{E}^{\rm IR} \vert^2$. As a consequence, the value of $\delta P_2 \left( \tau_{\rm end} ,\, \vec{k} \right)$ for a single realization coincides with (\ref{dP1-aniso}), with  $\vec{E}^{(0)}$ replaced by $ \vec{E}^{\rm IR} $,
 \begin{equation}
 \delta P_2 \left( \tau_{\rm end} ,\, \vec{k}  \right) \vert_{1 \, {\rm realization}}  \simeq \frac{24}{\epsilon} \, \frac{\vert \vec{E}^{\rm IR} \vert^2 }{V \left( \varphi \right) } \, N_k^2 {\cal P}^{(0)}   \sin^2  \theta_{{\hat k} , {\hat E }^{\rm IR}  }   
\label{dP2-aniso}
\end{equation}

Computing the theoretical expectation for this contribution amounts in replacing 
$ \sin^2  \theta_{{\hat k} , {\hat E }^{\rm IR}  }   \, E^{{\rm IR} 2} $ with $\frac{2}{3} \, \sigma^2_{ \vec{E}^{\rm IR},N}  $. The resulting expression coincides with (\ref{dP2-theory}).

We note that taking the classical IR value for one propagator is equivalent to taking the classical IR value for one of the two vector fields in the interaction hamiltonian $H_{{\rm int},2}$ given in eq. (\ref{Hint}). This interaction term then becomes identical to  $H_{{\rm int},1}$ apart from the fact that $\vec{E}^{(0)}$ is replaced by $\vec{E}^{\rm IR}$. The two expressions than become a unique vertex in terms of the vector  $ \vec{E}_{\rm classical } =  \vec{E}^{(0)} +   \vec{E}^{\rm IR} $. This is the expected physical result, since the  CMB modes ``measure'' the sum  $ \vec{E}_{\rm classical }  $ and cannot distinguish the two components. Even at the diagrammatic level, taking the IR limit in the third diagram of  Figure \ref{fig:zz} amounts in shrinking to zero one of the propagators, and one thus recovers the second diagram with a different  external vector in the vertex (clearly, all this discussion applies also to the bispectrum computation). Therefore, from the last two diagrams of 
 Figure \ref{fig:zz} we obtain 
 \begin{equation}
 \delta P \left( \tau_{\rm end} ,\, \vec{k}  \right) \simeq \frac{24}{\epsilon} \, \frac{E_{\rm classical}^2}{V \left( \varphi \right) } \, N_k^2 {\cal P}^{(0)}   \sin^2  \theta_{{\hat k} , {\hat E }_{\rm classical}  }   
\label{dP-aniso}
\end{equation}

\section{Anisotropic Bispectrum} \label{section:zzz}

\begin{widetext}
\begin{figure*}[ht!]
\centerline{
\includegraphics[width=0.6\textwidth,angle=0]{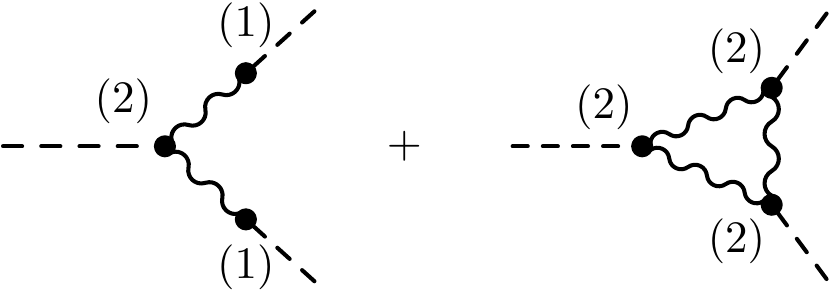}
}
\caption{Leading diagrams for $\langle \zeta^3 \rangle$, with the vertices labelled as in (\ref{Hint}).
}
\label{fig:zzz}\end{figure*}
\end{widetext}

The bispectrum of $\zeta$ is defined as 
\begin{equation}
{\cal B}_\zeta \left( \tau ,\, \vec{k}_1 ,\, \vec{k}_2 ,\, \vec{k}_3 \right) 
\delta^{(3)} \left( \sum_i \vec{k}_i \right)  = \left\langle {\hat \zeta}_{\vec{k}_1} \,  {\hat \zeta}_{\vec{k}_2} \,  {\hat \zeta}_{\vec{k}_3} \left( \tau \right)  \right\rangle
\end{equation}

The diagrams  shown in Figure \ref{fig:zzz} give  the  dominant contributions to the bispectrum. The computation is presented in the following Subsections.

\subsection{Tree level contributions} \label{subsection:zzz-tree}

We evaluate the contribution of the first diagram  ${\cal B}_1$ as we did for $\delta P_1$ in the previous Subsection. We start from
\begin{eqnarray}
\left\langle {\hat \zeta}_{\vec{k}_1} \,  {\hat \zeta}_{\vec{k}_2} \,  {\hat \zeta}_{\vec{k}_3} \left( \tau \right) 
 \right\rangle_{1} & = & i \int d \tau_1 d \tau_2 d \tau_3 \, 
\big\langle \, \big[ \big[ \big[ 
 {\hat \zeta}^{(0)}_{\vec{k}_1} \,  {\hat \zeta}^{(0)}_{\vec{k}_2} \,  {\hat \zeta}^{(0)}_{\vec{k}_3} \left( \tau \right) ,\, \nonumber\\
& & \!\!\!\!\!\!\!\!  \!\!\!\!\!\!\!\!  \!\!\!\!\!\!\!\!  \!\!\!\!
 {\cal H}_{{\rm int},2} \left( \tau_1 \right) \big] ,\,  {\cal H}_{{\rm int},1} \left( \tau_2 \right) \big] ,\, 
 {\cal H}_{{\rm int},1} \left( \tau_3 \right) \big] \big\rangle + \dots
\end{eqnarray}
where the dots denote two additional  terms obtained by permuting the position of  $ {\cal H}_{{\rm int},2} $. Performing the commutators between the ${\hat \zeta}^{(0)}$ fields results into
\begin{eqnarray}
& & 
\left\langle {\hat \zeta}_{\vec{k}_1} \,  {\hat \zeta}_{\vec{k}_2} \,  {\hat \zeta}_{\vec{k}_3} \left( \tau \right) 
 \right\rangle_{1}  \simeq \frac{4 \, E_x^{(0) 2}}{27 \epsilon^3 H^6 M_p^6} \prod_{i=1}^3 \int_{\tau_{\rm min}}^\tau
  d \tau_i \frac{\tau^3-\tau_i^3}{\tau_i^4}        \nonumber\\
& &  \quad\quad  \quad
 \int \frac{d^3 p}{\left( 2 \pi \right)^{3/2}} 
\Big\langle
\delta E_x \left( \tau_1 ,\, \vec{k}_1 \right) \, \delta E_x \left( \tau_2 ,\, \vec{k}_2  \right) 
\nonumber\\
& & \quad\quad\quad  \quad\quad \quad
 \delta \vec{E} \left( \tau_3 ,\, \vec{p} \right) \cdot \delta \vec{E} \left( \tau_3 ,\, \vec{k}_3 - \vec{p} \right) + \dots \Big\rangle
\label{partial-zzz1}
 \end{eqnarray}
where dots denote two additional terms obtained by permuting  $\vec{k}_3$ with the other two momenta. As for the diagrams evaluated in the previous Subsection, the time integrals are dominated by the times for which the mode functions are classical and do not oscillate. For the term that is explicitly written in  (\ref{partial-zzz1}) this gives the three lower limits  $\tau_1 >  - \frac{1}{k_1} $,  $\tau_2 > -\frac{1}{k_2}$, and $\tau_3 >  {\rm Max} \, \left[ - \frac{1}{k_1} ,\, - \frac{1}{k_2} \right] $ (the limit on $\tau_3$ is most easily seen after taking the expectation value). Taking the expectation value and performing the time integral, we obtain
\begin{eqnarray}
& & 
 {\cal B}_{1} \left( \tau_{\rm end}  ,\, \vec{k}_i \right)  \simeq \frac{ 288 \, \sqrt{2} \, \pi^{5/2} }{ \epsilon } \, \frac{E_x^{(0) 2}}{V \left( \varphi \right)} \, {\cal P}^{(0) 2}  \nonumber\\
 & &   \times \left\{ \frac{   N_{k_1} \,  N_{k_2} \, {\rm Min} \left[  N_{k_1} ,\,  N_{k_2} \right]  }{ k_1^3 \, k_2^3 } \,  {\cal C}_{{\hat k}_1 ,\, {\hat k}_2 , {\hat E}^{(0)} }    +  2 \; {\rm permutations}
 \right\} \nonumber\\
  \label{B-aniso1}
 \end{eqnarray}
where we have defined
\begin{eqnarray}
    {\cal C}_{{\hat k}_1 ,\, {\hat k}_2 , {\hat V} }    & \equiv &    1 - \cos^2 \theta_{ {\hat k}_1 ,\, {\hat V} } - \cos^2 \theta_{ {\hat k}_2 ,\, {\hat V} } \nonumber\\ 
      & & \quad\quad  +   \cos \theta_{ {\hat k}_1 ,\, {\hat V} } \cos \theta_{ {\hat k}_2 ,\, {\hat V} } \cos \theta_{ {\hat k}_1 ,\, {\hat k}_2 } 
\label{shape-aniso}
 \end{eqnarray}

\subsection{Including the loop contribution} \label{subsection:zzz-loop}

For the second diagram  in Figure \ref{fig:zzz} we obtain 
\begin{eqnarray}
& & \left\langle {\hat \zeta}_{\vec{k}_1} \,  {\hat \zeta}_{\vec{k}_2} \,  {\hat \zeta}_{\vec{k}_3} \left( \tau_{\rm end}  \right) 
 \right\rangle_{2}  \simeq   \frac{1}{27 \epsilon^3 H^6 M_p^6} \nonumber\\
 & & \left\langle \prod_{i=1}^3   \int \frac{d^3 p_i}{\left( 2 \pi \right)^{3/2}}  \int d \tau_i \frac{\tau^3 - \tau_i^3}{\tau_i^4} 
   \delta \vec{E} \left( \tau_i , \vec{p}_i \right) \cdot \delta \vec{E} \left( \tau_i , \vec{k}_i - \vec{p}_i \right) \right\rangle \nonumber\\
\label{partial-zzz2}
 \end{eqnarray}
where again we restrict the time integrals to when the modes are outside the horizon. The full correlator gives
\begin{eqnarray}
& &   \left\langle {\hat \zeta}_{\vec{k}_1} \,  {\hat \zeta}_{\vec{k}_2} \,  {\hat \zeta}_{\vec{k}_3} \left( \tau_{\rm end}  \right) 
 \right\rangle_{2}   \vert_{\rm theory}   \simeq   \frac{27 H^6 \delta^{(3)} \left( \sum_i \vec{k}_i \right)}{\epsilon^3 M_p^6}
\int \frac{d^3 p}{\left( 2 \pi \right)^{9/2}}  \nonumber\\
& & 
 \prod_{i=1}^3 \int d \tau_i \frac{\tau^3 - \tau_i^3  }{ \tau_i^4 }
\frac{Q_{ki} \left[ {\hat p} \right] \, Q_{ij} \left[ \widehat{p-k_1} \right] \,  Q_{jk} \left[ \widehat{p+k_3} \right]  
}{p^3 \, \vert \vec{p} - \vec{k}_1 \vert^3 \, \vert \vec{p} + \vec{k}_3 \vert^3} \nonumber\\
\end{eqnarray}  
where we have defined $Q_{ij} \left[ {\hat p} \right] \equiv  \delta_{ij} - {\hat p}_i \, {\hat p}_j $ and where the time integrals are restricted to $\tau_1 > {\rm Max } \left[ - \frac{1}{p} , - \frac{1}{\vert \vec{p} - \vec{k}_1 \vert} \right] $, 
 $\tau_2 > {\rm Max } \left[ - \frac{1}{\vert \vec{p} - \vec{k}_1 \vert} , - \frac{1}{\vert \vec{p} + \vec{k}_3 \vert} \right] $, 
and  $\tau_3 > {\rm Max } \left[ - \frac{1}{p} , - \frac{1}{\vert \vec{p} + \vec{k}_3 \vert} \right] $. 

The final momentum integral has three poles where it naively diverges logarithmically. The integral is regulated by the same argument used after (\ref{partial3-zz3}), which in the present case enforces $p ,\, \vert p - \vec{k}_1 \vert ,\, \vert \vec{p} + \vec{k}_3 \vert > - \frac{1}{\tau_{\rm in} } $. Performing the integrals gives
\begin{eqnarray}
& &  {\cal B}_{2} \left( \tau_{\rm end}  ,\, \vec{k}_i \right) \vert_{\rm theory}  \simeq  1152 \sqrt{2} \pi^{5/2} {\cal P}^{(0) 3} 
  \Big\{ N_{k_1} N_{k_2}  \nonumber\\
& & 
\quad \times  {\rm Min } \left[ N_{k_1} \left( N_{\rm tot} - N_{k_1} \right) ,  N_{k_2} \left( N_{\rm tot} - N_{k_2} \right) \right]  \, \frac{1 + \cos^2 \theta_{{\hat k}_1 {\hat k}_2}}{k_1^3 k_2^3} \nonumber\\
& & \quad\quad +  2 \; {\rm permutations} \Big\} 
  \label{B2-theory}
 \end{eqnarray}

As for (\ref{dP2-theory}), this result is accurate for $N_{\rm tot} \gg N_{k_i}$; if this is not the case, the factor $N_{\rm tot} - N_{k_i}$ should be replaced by a ${\rm O } \left( 1 \right)$ factor.  The result (\ref{B2-theory}) was first derived in  \cite{Barnaby:2012tk} using the Green function method.

As for $\delta P_2$, the result is dominated by the regime in which one propagator has an IR mode. Also in this case, the value for a single realization is obtained by replacing the quantum operator of the IR field entering in (\ref{partial-zzz2}) by a classical Fourier transform. We recover an expression formally identical to (\ref{partial-zzz1}), with   $\vec{E}^{(0)}$ replaced by $ \vec{E}^{\rm IR} $. Therefore, precisely as for the power spectrum, the total contribution of the two diagrams to the bispectrum is given by
\begin{eqnarray}
& & 
 {\cal B} \left( \tau_{\rm end}  ,\, \vec{k}_i \right)  \simeq \frac{ 288 \, \sqrt{2} \, \pi^{5/2} }{ \epsilon } \, \frac{\vert \vec{E}_{\rm classical} \vert^2}{V \left( \varphi \right)} \, {\cal P}^{(0) 2}  \nonumber\\
 & &   \times \left\{ \frac{   N_{k_1} \,  N_{k_2} \, {\rm Min} \left[  N_{k_1} ,\,  N_{k_2} \right]  }{ k_1^3 \, k_2^3 } \,  {\cal C}_{{\hat k}_1 ,\, {\hat k}_2 , {\hat E}^{\rm classical} }    +  2 \; {\rm permut.}
 \right\} \nonumber\\
  \label{B-aniso}
 \end{eqnarray}
where we recall that ${\cal C}$ is given in (\ref{shape-aniso}).

\section{Phenomenology}
\label{sec:phenomenology}

The total power spectrum is given by
\begin{equation}
P_\zeta = {\cal P}^{(0)} \left[ 1 +    \frac{24}{\epsilon} \, \frac{E_{\rm classical}^2}{V \left( \varphi \right) } \,  
\left( 1 - \cos^2  \theta_{{\hat k} , {\hat E }_{\rm classical} } \right)  \right]
\label{Pz-tot}
\end{equation}
corresponding to a negative $g_*$ parameter in (\ref{acw})
\begin{eqnarray}
g_* & = & - \frac{24}{\epsilon} \,  \frac{E_{\rm classical}^2}{V \left( \varphi \right) } \,  N_k^2
\Bigg/ \left[ 1 +  \frac{24}{\epsilon} \,  \frac{E_{\rm classical}^2}{V \left( \varphi \right) } \,  N_k^2 \right]
\nonumber\\
& \simeq &  - \frac{24}{\epsilon} \,  \frac{E_{\rm classical}^2}{V \left( \varphi \right) } \,  N_k^2
\end{eqnarray}
where the approximation is due to the fact that  $\vert g_* \vert$ is phenomenologically constrained to be $<< 1$.

Inverting this relation for the CMB modes, and denoting by $\rho_{E_{\rm cl}} \equiv \frac{E_{\rm classical}^2}{2}$ the  energy of the classical ``electric'' field present when the CMB modes left the horizon
\begin{equation}
 \frac{\rho_{E_{\rm cl}}  }{V \left( \varphi \right)} \simeq 5.8 \cdot 10^{-9} \, \frac{\epsilon}{0.01} \,  \frac{\vert g_* \vert_{\rm CMB} }{0.1} \, \left( \frac{60}{N_{\rm CMB} } \right)^2
\label{rhoE-V}
\end{equation}
and therefore we see that even a very subdominant vector field can produce an appreciably anisotropic power spectrum. Conversely, this indicates that the classical ``electric'' field must be extremely subdominant not to conflict with the phenomenological limits on the isotropy of the  power spectrum. 

We recall that, in any given realization, $\vec{E}_{\rm classical} = \vec{E}^{(0)} + \vec{E}^{\rm IR}$, where $ \vec{E}^{(0)} $ is the solution of the classical equations of motion of the model, and $ \vec{E}^{\rm IR} $ is drawn by a gaussian statistics of zero mean, and variance (\ref{variance-EIR})
\begin{equation}
\left\langle \vec{E}^{\rm IR} \;^2 \right\rangle_N = \frac{9 H^4}{2 \pi^2} \, N
\label{variance2}
\end{equation}
The variance grows with the number of e-folds of inflation.

It is instructive to evaluate the ratio
\begin{equation}
R_N \equiv \frac{ \rho_{E_{\rm cl}} }{ \rho_{{\delta E},N} } = \frac{\left\vert \vec{E}^{(0)}  + \vec{E}^{\rm IR} \right\vert^2}{\left\langle \vec{E}^{\rm IR} \;^2 \right\rangle_N }
\label{RN-def}
\end{equation}
at the moment that the CMB fluctuations left the horizon.  Combining (\ref{rhoE-V}) and (\ref{variance2}), and using the observed value   \cite{Komatsu:2010fb}  for the dominant term 
  ${\cal P}^{(0)} \simeq P_\zeta \simeq 2.5 \cdot 10^{-9} \,$,  we obtain
\begin{equation}
R_{N_{\rm tot} - N_{\rm CMB}} \simeq \frac{\vert g_* \vert_{\rm CMB}}{0.1} \, \left( \frac{60}{N_{\rm CMB}} \right)^2 \, \frac{37}{N_{\rm tot} - N_{\rm CMB}} 
\label{RN-result}
\end{equation}

The numerator of (\ref{RN-def}) is  the observed value, while the denominator  the theoretical variance. Therefore, a value  $ R_{N_{\rm tot} - N_{\rm CMB}} \geq 1$ should be naturally expected for this ratio. Indeed,  $  R_{N_{\rm tot} - N_{\rm CMB}} \ll 1 $ indicates either that the sum of the IR modes is unnaturally small in that realization (i.e. $
\vert \vec{E}^{(0)} \vert^2 ,\,  \vert \vec{E}^{\rm IR} \vert^2 \ll \left\langle \vec{E}^{\rm IR} \;^2 \right\rangle_{N_{\rm tot} - N_{\rm CMB}}$), or that it unnaturally cancels against $\vec{E}^{(0)}$. Assuming  a natural $R_{N_{\rm tot} - N_{\rm CMB}} \ga 1$ realization,~\footnote{Strictly speaking, eq. (\ref{rhoE-V}), and those derived from it, are only valid for $ \vert g_* \vert < 1$. A value of $N_{\rm tot} - N_{\rm CMB}$ that results in a $ \vert g_* \vert > 1$ in (\ref{RN-result}) should be associated to a $\vert g_* \vert = {\rm O } \left( 1 \right)$ in the natural $R_{N_{\rm tot} - N_{\rm CMB}} \ga 1$ regime.}
\begin{equation}
\vert g_* \vert_{\rm CMB} \ga {\rm Min } \left[  0.1 \, \left( \frac{N_{\rm CMB}}{60}  \right)^2 \, \frac{N_{\rm tot} - N_{\rm CMB}}{37} ,\, 1 \right] 
\label{gstar}
\end{equation}
In generic slow roll inflationary potentials , the total duration of inflation exceeds (by much) the minimal $N_{\rm CMB}$ amount. One naturally finds an order one anisotropy in these models.

Another important conclusion can be drawn from (\ref{RN-result}).  As it is exponentially unlikely to have $\vert  \vec{E}^{\rm IR} \vert^2 $ much greater than its variance, a value $ R_{N_{\rm tot} - N_{\rm CMB}} \gg 1$ implies that $\vert \vec{E}^{(0)} \vert$ is much greater than the sum of the IR modes. Only in this case  the solution of the classical equations of motion provides an accurate value for the classical vector field that affects the observables curvature perturbations (as it is typically assumed in many works). A value  $ R_{N_{\rm tot} - N_{\rm CMB}}   \simeq 1 $  indicates instead that the  contribution of the IR modes is no longer negligible, and that the solution of the classical equations of motion is unstable under quantum fluctuations (more appropriately, under the sum of the quantum fluctuations that have become classical). We see from (\ref{RN-result}) that 
 $ R_{N_{\rm tot} - N_{\rm CMB}} \gg 1$ can be obtained only for $N_{\rm tot} - N_{\rm CMB} \ll 37$. 

Therefore, this mechanism can result in a $ \vert g_* \vert_{\rm CMB} \simeq 0.1$ anisotropy for either  a tuned short duration of zinflation or   an unnaturally small $\vec{E}_{\rm classical}$. Assuming that this is the case, we can obtain a firm prediction from the anisotropy, the shape, and the magnitude of the non-gaussianity in the model. The bispectrum from this mechanism is given in eq. (\ref{B-aniso}).  To quantify whether this bispectrum is of observable magnitude, we notice that it is enhanced in the squeezed limit precisely as the local template. We therefore insert  (\ref{B-aniso}) into the relation
\begin{equation}
{\cal B}_\zeta  \left( \tau_{\rm end} ,\, \vec{k}_i \right) \equiv \frac{3}{10} \left( 2 \pi \right)^{5/2} f_{\rm NL} P_\zeta \left( k \right)^2 
\frac{\sum_i k^3}{\prod_i k_i^3}
\label{fNL}
\end{equation}
which, for the local bispectrum template,    results in a constant  $f_{\rm NL}^{\rm local}$ with the correct normalization (the non conventional factor of ``$2 \pi$-dependence'' is due to our choice (\ref{zeta-0}) for the Fourier transform). In the squeezed limit, we obtain
\begin{equation}
f_{\rm NL} \simeq \frac{ 480 }{ \epsilon } \, \frac{{\cal P}^{(0) 2}}{P_\zeta^2} \,  \frac{\rho_{E_{\rm cl}}}{V \left( \varphi \right)} \, 
     N_{k_1} N_{k_2}^2 \,  {\cal C}_{{\hat k}_1 ,\, {\hat k}_2 , {\hat V} }  \;\;\;,\;\;\; k_1 \ll k_2 \simeq k_3
\end{equation}
where ${\cal C}$ is given in (\ref{shape-aniso}). Recalling that ${\cal P}^{(0)}$ dominates the power spectrum, using 
(\ref{rhoE-V}), and simply denoting $N_{k_1} \simeq N_{k_2} \equiv N_{\rm CMB}$ (assuming that the momenta are not too hierarchical; this is certainly the case in the CMB analysis), we arrive to the final estimate
\begin{equation}
f_{\rm NL} \simeq 26 \, \frac{\vert g_{*} \vert_{\rm CMB}}{0.1} \,  \frac{N_{\rm CMB}}{60} \, \frac{  {\cal C}_{{\hat k}_1 ,\, {\hat k}_2 , {\hat V} } }{ 4/9 } \;\;\;,\;\;\; k_1 \ll k_2 \simeq k_3
\label{estimate-fnl}
\end{equation}
 where  the value $4/9$ has been inserted in the last factor so that this factor averages to one if we average over all directions of $\vec{k}_1$ and $\vec{k}_2 \simeq \vec{k}_3$. This result holds independently of the type of inflationary potential considered. In the estimate (\ref{estimate-fnl}) we have averaged over all directions, with the assumption that this is  what is done when extracting the value of $f_{\rm NL}$ from the full-sky data. The estimate (\ref{estimate-fnl}) indicates that a  $\vert g_* \vert$ in the $0.01-0.1$ range from this mechanism can likely be  associated to a detectable bispectrum. A  detection of a non- vanishing $f_{\rm NL}$ of the local shape will motivate a more detailed analysis, where the full shape and the anisotropy of  (\ref{B-aniso}) are retained.

\section{Applications}
\label{sec:apps}

In the following subsections we discuss how our findings apply to three well studied models in which vector fields are employed to either break statistical isotropy or to generate primordial magnetic fields.

\subsection{Anisotropic inflation}
\label{subsec:Watanabe}

Ref.  \cite{Watanabe:2009ct} studied anisotropic inflation  from  the model  (\ref{example-VI}). It was shown that the classical equation of the system admit an attractor solution characterized by the geometry  (\ref{bianchi}) and by
\begin{eqnarray}
E_x^{(0)} \simeq \sqrt{3 \, \epsilon \left( c - 1 \right)} \, H \, M_p 
\label{E-W--back}
\end{eqnarray}

This values corresponds to  \cite{Watanabe:2009ct} 
\begin{equation}
\frac{\Delta H}{H} \simeq \frac{2 \, \rho_{E^{(0)}}}{V \left( \varphi \right)}   \simeq  \frac{\left( c - 1 \right) \, \epsilon}{ c^2} 
\label{dH-H}
\end{equation}

The power spectrum in this model was computed in \cite{Dulaney:2010sq,Gumrukcuoglu:2010yc,Watanabe:2010fh} that however disregarded the contribution of $\vec{E}^{\rm IR}$. Our computation reproduces their result (which confirms the validity of our computational scheme and the accuracy of our approximations) in this limit. Moreover, we have computed for the first time the bispectrum of this model, using the same computational scheme used to compute the power spectrum.

The contribution of $\vec{E}^{(0)}$ to the anisotropy of the power spectrum forces $c-1 \la 10^{-6}$  \cite{Dulaney:2010sq,Gumrukcuoglu:2010yc,Watanabe:2010fh}. The implicit assumption in this result is that choosing a $c-1 \rightarrow 0$ in
the lagrangian would result in a $g_* \rightarrow 0$ parameter. However, taking also $\vec{E}^{\rm IR}$ into consideration, leads to the general conclusions on the  natural amount of $g_*$ and on the instability of (\ref{E-W--back}) that we have presented in the previous section.

\subsection{Waterfall mechanism}

Ref. \cite{Yokoyama:2008xw} embedded the mechanism (\ref{L-ratra}) in hybrid inflation, through the lagrangian
\begin{eqnarray}
{\cal L} & = & - \frac{I^2 \left( \varphi \right) F^2}{4} - \frac{1}{2} \left( \partial \varphi \right)^2  - \frac{1}{2} \left( \partial \chi \right)^2 
- V \nonumber\\
V & = & \frac{\lambda}{4} \left( \chi^2 - v^2 \right)^2 + \frac{1}{2} g^2 \varphi^2 \chi^2 + \frac{1}{2} m^2 \varphi^2 + \frac{1}{2} h^2 A^\mu A_\mu \chi^2 \nonumber\\
\label{YS}
\end{eqnarray}

The additional  coupling of the vector $A_\mu$ to the waterfall field  $\chi$ provides a contribution to $\zeta$ through the mechanism of modulated perturbations \cite{modulated}. Specifically, the value of the vector field contributes to determine the end of inflation, that happens when
\begin{equation}
m_\chi^2 = - \lambda v^2 + g^2 \varphi^2 + h^2 A_\mu A^\mu = 0
\end{equation}
so that perturbations $\delta A_\mu$ get converted into $\zeta$ at the end of inflation. This results in the additional contributions
 \cite{Yokoyama:2008xw} 
\begin{equation}
\Delta g_* \sim -  \frac{h^4 \vert A \vert^2}{g^4 \varphi_e^2}  \;\;\;,\;\;\;
\Delta f_{\rm NL} \sim \eta_e \left( \frac{h^2}{g^2} \, \Delta g_* - \Delta g_*^2 \right)
\label{res-YS}
\end{equation}
where $\varphi_e$ and $\eta_e$ are, respectively,  the value of the inflaton and of the slow roll parameter $\eta$  at the end of inflation, and where we have normalized the coupling in 
 (\ref{L-ratra}) to $ \langle  I \rangle  \equiv 1 $ at the end of inflation.\footnote{Ref. \cite{Lyth:2012vn} studied the additional contributions to $\zeta$ from the waterfall coupling in the case in which the vev of the vector field varies with time. For a discussion of the effects from a time varying inflaton coupling in modulated reheating see~\cite{Matarrese:2003tk}. }

These  additional contributions to the anisotropy obviously vanish in the limit in which the vector field is not coupled to the waterfall field ($h \rightarrow 0$). On the contrary,   (\ref{gstar}) and (\ref{estimate-fnl}) are  general results unavoidably present when the function $I$ in (\ref{YS}) is arranged to provide a constant vector field (which is also necessary to have non-negligible values in (\ref{res-YS})). These unavoidable contributions have not been included in the many works that studied  \cite{Yokoyama:2008xw}. Given that  (\ref{gstar}) nearly saturates the phenomenological limit, the contribution (\ref{res-YS}) to $g_*$ is at most comparable to  (\ref{gstar}). 
On the contrary, the additional coupling $h$ present in 
(\ref{YS}) can allow for a greater non-gaussianity than  (\ref{estimate-fnl}). Indeed, in the minimal model (\ref{L-ratra}), the parameter $f_{\rm NL}$ is uniquely related to $g_*$ through  (\ref{estimate-fnl}), and this strict relation increases the predictivity of the mechanism. For the model (\ref{YS}), non-gaussianity can be enhanced for $h \gg g$. The shape of the contribution to the bispectrum resulting from the waterfall coupling  \cite{Yokoyama:2008xw}  is in general different from the shape of  (\ref{B-aniso}); the two shapes coincide  for $h \gg g$.

\subsection{Magnetogenesis}

Using the ``electric'' $ \leftrightarrow $ ``magnetic'' duality that we have mentioned \cite{Giovannini:2009xa}, the results for the perturbations that we have obtained are also valid in the case of $\langle I \rangle \propto a^2$, which results in a scale invariant ``magnetic'' field \cite{Ratra:1991bn}. In fact, the two results (\ref{dP2-theory}) and (\ref{B2-theory}) have been first obtained in \cite{Barnaby:2012tk} for  $\langle I \rangle \propto a^{2}$. Although the mechanism  (\ref{L-ratra}) was first introduced with this motivation, the magnetogenesis application suffers a strong coupling problem  \cite{Demozzi:2009fu}.  Indeed, with  (\ref{L-ratra}) 
 the fine structure constant scales as $\alpha \propto \langle I \rangle^{-2}$. The choice  $\langle I \rangle \propto a^2$ corresponds to a fast decreasing $\alpha \propto a^{-4}$ during inflation. Assuming that $\alpha$ reaches the present value at the end of inflation implies  $\alpha \gg 1$ during inflation, and   a quantum field theory of electromagnetism that is (at the very least) out of computational control  \cite{Demozzi:2009fu}. Alternatively, for $\alpha \la 1$ at the start of inflation, one obtains $\alpha \sim {\rm e}^{-4N_{\rm tot}}$ at the end of inflation, and requiring that $\alpha$ grows back to its present value before  big-bang nucleosynthesis does not appear feasible  \cite{Barnaby:2012tk}.~\footnote{A scale invariant magnetic field is also obtained at the start of inflation for $\langle I \rangle \propto a^{-3}$. However, too much energy gets stored in the electric field in this case  \cite{Demozzi:2009fu}, and  the system enters in a strong backreaction regime in which a too small magnetic field is produced \cite{Kanno:2009ei}.}

Clearly, the strong coupling problem is a problem of the magnetogenesis application and not of the mechanism (\ref{L-ratra}), as one can always assume that a generic U(1) field  is sufficiently weakly coupled to other fields.  For a growing $\langle I \rangle$ 
the coupling decreases during inflation, and one can  assume that the vector field has couplings $\la 1$ at the beginning of inflation. For a decreasing $\langle I \rangle$ the coupling grows during inflation, and one can require a coupling $\la 1$ when inflation ends. In principle one may hope that a  field with such properties is  produced during inflation by this mechanism, and it is then (partially) converted to the magnetic one through some coupling. Some attempts so far in this direction have been unsuccessful \cite{Barnaby:2012tk}, but a general study remains to be done. If this, or some other idea
\cite{ratra-pert}, can circumvent the strong coupling problem, our findings provide a signature that may be correlated with the magnetic field.

\section{Conclusions}
\label{sec:conclusions}

There are two questions associated to the attempt of reconciling  a non vanishing anisotropy $g_*$ with a model of inflation. The first one is ``what are the other signatures that would accompany the measured value of $g_*$ ?"; the second one is ``what is the natural value of $g_*$ in the model ?". A first difficulty that is encountered in answering these questions is that anisotropic hairs are typically erased by inflation \cite{Wald:1983ky,Maleknejad:2012as}, so that one has to design specific models that preserve the anisotropy. Vector fields appear as the simplest possibility, as their vev breaks isotropy. Several mechanisms have been designed to prevent the  rapidly erosion of the vector vev by the inflationary expansion. As shown in \cite{hcp}, however,  many of them (for instance, the use of a potential as in \cite{Ford:1989me}, of a lagrange multiplier as in  \cite{Ackerman:2007nb}, or of a coupling to the curvature as in \cite{Golovnev:2008cf} and as in analogous models of vector curvaton) have ghosts. The use of (\ref{L-ratra}) for preserving the vector vev  \cite{Watanabe:2009ct,Yokoyama:2008xw} has resulted in a rather exceptional mechanism where the above two questions can be formulated.

With the mechanism  (\ref{L-ratra}), the answer to the first question appears to be very positive. The anisotropy in the power spectrum is correlated with a characteristic and very likely detectable bispectrum. The bispectrum, whose full shape is given in (\ref{B-aniso}), is enhanced as the local one in the squeezed limit, where it has an effective local parameter
(see (\ref{estimate-fnl}) for  more details)
\begin{equation}
f_{\rm NL} \simeq 26 \, \frac{\vert g_{*} \vert_{\rm CMB}}{0.1} 
\label{fNL-conclusions}
\end{equation}
A larger bispectrum can be obtained if the vector field has additional interactions with the inflationary sector, as in the waterfall mechanism of \cite{Yokoyama:2008xw}. However these additional interactions are not needed, and the result 
(\ref{fNL-conclusions}) is the direct contribution that comes from the interaction  (\ref{L-ratra}). Therefore, this is the result in the most minimal and predictive implementation of  (\ref{L-ratra}). We see that an anisotropy in the power spectrum at the $1\%-10\%$ level, could be associated with a detectable bispectrum.

For the second question, one needs to take into account that, for a mechanism that results in a scale invariant  field (as in the implementations  \cite{Watanabe:2009ct,Yokoyama:2008xw}), a classical background value of this field is unavoidably generated by the sum of the modes that have left the horizon during inflation \cite{Linde:2007fr}. The crucial point for our discussion is that a background vector field breaks isotropy. It is unnatural to require that the anisotropy experienced by the CMB modes when they leave the horizon is smaller than the theoretical expectation value $\left\langle \delta \vec{E}^2 \right\rangle \sim H^4 \, \left( N_{\rm tot} - N_{\rm CMB} \right)$, where $H$ is the Hubble rate during inflation, and $ N_{\rm tot} - N_{\rm CMB} $ is the number of e-folds of inflation that took place before the CMB modes are generated, and during which the classical background of long-wavelength modes experienced by the CMB modes has accumulated. Taking this into account, we find that the natural value for the anisotropy is (see (\ref{gstar}) for  more details)
\begin{equation}
\vert g_* \vert_{\rm CMB} \ga {\rm Min } \left[  0.1 \,  \frac{N_{\rm tot} - N_{\rm CMB}}{37} ,\, 1 \right] 
\label{gstar-conclusions}
\end{equation}
which can easily overcome the phenomenological bounds (particularly with the improvement expected from Planck \cite{Pullen:2007tu,MA}). We note that this saturation takes place for a very subdominant vector field, with very negligible backreaction on the background inflaton evolution. However, as already obtained in  \cite{Dulaney:2010sq,Gumrukcuoglu:2010yc,Watanabe:2010fh}, even a very subdominant vector field can result in a substantial anisotropy parameter.

Remarkably, these considerations apply also to the magnetogenesis applications of (\ref{L-ratra}). The role of the background magnetic field has been so far disregarded in the study of the cosmological perturbations obtained in that context. However, the result (\ref{gstar-conclusions}) applies also to this case. We stress that the magnetogenesis application suffers from a serious strong coupling problem  \cite{Demozzi:2009fu}. If this problem can be solved, a magnetic field through this mechanism would be naturally correlated with an observable anisotropy of the perturbations.

Analogously, an anisotropic signal should also be expected in models where a triad of vectors is arranged to produce isotropic expansion or isotropic perturbations \cite{triad}, since the long-wavelength background values of the different vectors fields will in general be different.

Although our study has been limited to the mechanism (\ref{L-ratra}), one may expect that an anisotropic background  field and a large anisotropy of the perturbations may be a general outcome of all models that sustain higher  than $0$ spin fields during inflation. For instance, it would be interesting to study the natural level of anisotropy to be expected in the mechanism of \cite{Dimopoulos:2009am} (for which  statistical isotropy has been obtained for very specific choices of the kinetic and the mass function \cite{Namba:2012gg}) or in the case of inflation or magnetic fields from $p$-forms \cite{pforms}.

\vskip.25cm
\noindent{\bf Acknowledgements:} 
The work of N.B. and S.M. was partially supported by the ASI/INAF Agreement 
I/072/09/0 for the Planck LFI Activity of Phase E2. N.B, S.M. and A.R.  were also supported by the PRIN 2009 project "La Ricerca di non-Gaussianit{\' a} Primordiale". The work of  M.P. was partially supported  by DOE grant DE-FG02-94ER-40823 at the University of Minnesota. MP would like to thank the University of Padua for their friendly hospitality and for partial support during his sabbatical leave.


\begin{thebibliography}{99}



\bibitem{Neronov:1900zz} 
  A.~Neronov and I.~Vovk,
  Science {\bf 328}, 73 (2010)
  [arXiv:1006.3504 [astro-ph.HE]].

    
\bibitem{Groeneboom:2008fz} 
  N.~E.~Groeneboom and H.~K.~Eriksen,
  Astrophys.\ J.\  {\bf 690}, 1807 (2009)
  [arXiv:0807.2242 [astro-ph]].
     
\bibitem{Hanson:2009gu}
  D.~Hanson and A.~Lewis,
  Phys.\ Rev.\ D {\bf 80} (2009) 063004
  [arXiv:0908.0963 [astro-ph.CO]].
  
\bibitem{Groeneboom:2009cb} 
  N.~E.~Groeneboom, L.~Ackerman, I.~K.~Wehus and H.~K.~Eriksen,
  Astrophys.\ J.\  {\bf 722}, 452 (2010)
  [arXiv:0911.0150 [astro-ph.CO]].
  
\bibitem{Komatsu:2010fb} 
  E.~Komatsu {\it et al.}  [WMAP Collaboration],
  Astrophys.\ J.\ Suppl.\  {\bf 192}, 18 (2011)
  [arXiv:1001.4538 [astro-ph.CO]].
    
   
\bibitem{Ackerman:2007nb} 
  L.~Ackerman, S.~M.~Carroll and M.~B.~Wise,
  Phys.\ Rev.\ D {\bf 75}, 083502 (2007)
  [Erratum-ibid.\ D {\bf 80}, 069901 (2009)]
  [astro-ph/0701357].


  
\bibitem{Hanson:2010gu} 
  D.~Hanson, A.~Lewis and A.~Challinor,
  Phys.\ Rev.\ D {\bf 81}, 103003 (2010)
  [arXiv:1003.0198 [astro-ph.CO]].
 
\bibitem{Pullen:2007tu}
  A.~R.~Pullen and M.~Kamionkowski,
  Phys.\ Rev.\ D {\bf 76} (2007) 103529
  [arXiv:0709.1144 [astro-ph]].
        
   \bibitem{MA}
  Y.~Z.~Ma, G.~Efstathiou and A.~Challinor,
  Phys.\ Rev.\  D {\bf 83}, 083005 (2011)
  [arXiv:1102.4961 [astro-ph.CO]].

\bibitem{Pullen:2010zy} 
  A.~R.~Pullen and C.~M.~Hirata,
  JCAP {\bf 1005}, 027 (2010)
  [arXiv:1003.0673 [astro-ph.CO]].


\bibitem{Gumrukcuoglu:2006xj} 
A.~E.~Gumrukcuoglu, C.~R.~Contaldi and M.~Peloso,  astro-ph/0608405,  
Proceeding of the Ò11th Marcel Grossmann Meeting On General RelativityÓ Ed. H. Kleinert, R.T. Jantzen and R. Ruffini. Hackensack, World Scientific, 2008;
      
\bibitem{Wald:1983ky} 
  R.~M.~Wald,
  Phys.\ Rev.\ D {\bf 28}, 2118 (1983).
  
\bibitem{Maleknejad:2012as} 
  A.~Maleknejad and M.~M.~Sheikh-Jabbari,
  Phys.\ Rev.\ D {\bf 85}, 123508 (2012)
  [arXiv:1203.0219 [hep-th]].

\bibitem{Ford:1989me} 
  L.~H.~Ford,
  Phys.\ Rev.\ D {\bf 40}, 967 (1989).
  
\bibitem{Golovnev:2008cf} 
  A.~Golovnev, V.~Mukhanov and V.~Vanchurin,
  JCAP {\bf 0806}, 009 (2008)
  [arXiv:0802.2068 [astro-ph]].
  
\bibitem{Kanno:2008gn} 
  S.~Kanno, M.~Kimura, J.~Soda and S.~Yokoyama,
  JCAP {\bf 0808}, 034 (2008)
  [arXiv:0806.2422 [hep-ph]].
  
\bibitem{Turner:1987bw} 
  M.~S.~Turner and L.~M.~Widrow,
  Phys.\ Rev.\ D {\bf 37}, 2743 (1988).

\bibitem{hcp}
  B.~Himmetoglu, C.~R.~Contaldi and M.~Peloso,
  Phys.\ Rev.\ Lett.\  {\bf 102}, 111301 (2009)
  [arXiv:0809.2779 [astro-ph]];
 %
  B.~Himmetoglu, C.~R.~Contaldi and M.~Peloso,
  Phys.\ Rev.\ D {\bf 79}, 063517 (2009)
  [arXiv:0812.1231 [astro-ph]];
  %
  B.~Himmetoglu, C.~R.~Contaldi and M.~Peloso,
  Phys.\ Rev.\ D {\bf 80}, 123530 (2009)
  [arXiv:0909.3524 [astro-ph.CO]].

  
\bibitem{Watanabe:2009ct} 
  M.~-a.~Watanabe, S.~Kanno and J.~Soda,
  Phys.\ Rev.\ Lett.\  {\bf 102}, 191302 (2009)
  [arXiv:0902.2833 [hep-th]].
 
 \bibitem{aniso-fAA} 
  R.~Emami, H.~Firouzjahi, S.~M.~Sadegh Movahed and M.~Zarei,
  JCAP {\bf 1102}, 005 (2011)
  [arXiv:1010.5495 [astro-ph.CO]];
 %
  S.~Kanno, J.~Soda and M.~-a.~Watanabe,
  JCAP {\bf 1012}, 024 (2010)
  [arXiv:1010.5307 [hep-th]];
 %
  K.~Murata and J.~Soda,
  JCAP {\bf 1106}, 037 (2011)
  [arXiv:1103.6164 [hep-th]];
  %
  K.~Yamamoto, M.~-a.~Watanabe and J.~Soda,
  Class.\ Quant.\ Grav.\  {\bf 29}, 145008 (2012)
  [arXiv:1201.5309 [hep-th]];
  %
  M.~Thorsrud, D.~F.~Mota and S.~Hervik,
  arXiv:1205.6261 [hep-th].
 
 
 

\bibitem{Dimopoulos:2006ms} 
  K.~Dimopoulos,
  Phys.\ Rev.\ D {\bf 74}, 083502 (2006)
  [hep-ph/0607229].
           
\bibitem{Dimopoulos:2009am} 
  K.~Dimopoulos, M.~Karciauskas and J.~M.~Wagstaff,
  Phys.\ Rev.\ D {\bf 81}, 023522 (2010)
  [arXiv:0907.1838 [hep-ph]].
      
\bibitem{Afm-study}
  K.~Dimopoulos, M.~Karciauskas and J.~M.~Wagstaff,
  Phys.\ Lett.\ B {\bf 683}, 298 (2010)
  [arXiv:0909.0475 [hep-ph]];
%
  J.~M.~Wagstaff and K.~Dimopoulos,
  Phys.\ Rev.\ D {\bf 83}, 023523 (2011)
  [arXiv:1011.2517 [hep-ph]];
%
  K.~Dimopoulos,
  Int.\ J.\ Mod.\ Phys.\ D {\bf 21}, 1250023 (2012)
  [Erratum-ibid.\ D {\bf 21}, 1292003 (2012)]
  [arXiv:1107.2779 [hep-ph]];
%
  K.~Dimopoulos, D.~Wills and I.~Zavala,
  arXiv:1108.4424 [hep-th];
%
  K.~Dimopoulos and M.~Karciauskas,
  JHEP {\bf 1206}, 040 (2012)
  [arXiv:1203.0230 [hep-ph]].
    
\bibitem{Namba:2012gg} 
  R.~Namba,
  arXiv:1207.5547 [astro-ph.CO].

\bibitem{Ratra:1991bn} 
  B.~Ratra,
  Astrophys.\ J.\  {\bf 391}, L1 (1992).
        
\bibitem{Demozzi:2009fu} 
  V.~Demozzi, V.~Mukhanov and H.~Rubinstein,
  JCAP {\bf 0908}, 025 (2009)
  [arXiv:0907.1030 [astro-ph.CO]].
      
\bibitem{Giovannini:2009xa} 
  M.~Giovannini,
  JCAP {\bf 1004}, 003 (2010)
  [arXiv:0911.0896 [astro-ph.CO]].
  
\bibitem{Yokoyama:2008xw} 
  S.~Yokoyama and J.~Soda,
  JCAP {\bf 0808}, 005 (2008)
  [arXiv:0805.4265 [astro-ph]].
 

  
  
  
\bibitem{Karciauskas:2008bc} 
  M.~Karciauskas, K.~Dimopoulos and D.~H.~Lyth,
  Phys.\ Rev.\ D {\bf 80}, 023509 (2009)
  [Erratum-ibid.\ D {\bf 85}, 069905 (2012)]
  [arXiv:0812.0264 [astro-ph]].
    

  
\bibitem{Karciauskas:2011fp} 
  M.~Karciauskas,
  JCAP {\bf 1201}, 014 (2012)
  [arXiv:1104.3629 [astro-ph.CO]].
  
  
\bibitem{Lyth:2012br} 
  D.~H.~Lyth and M.~Karciauskas,
  arXiv:1204.6619 [astro-ph.CO].






\bibitem{Lyth:2012vn} 
  D.~H.~Lyth and M.~Karciauskas,
  arXiv:1209.4266 [astro-ph.CO].
    
\bibitem{Emami:2011yi} 
  R.~Emami and H.~Firouzjahi,
  JCAP {\bf 1201}, 022 (2012)
  [arXiv:1111.1919 [astro-ph.CO]].
         
  
  
  
  
  
  
\bibitem{Himmetoglu:2009mk} 
  B.~Himmetoglu,
  JCAP {\bf 1003}, 023 (2010)
  [arXiv:0910.3235 [astro-ph.CO]].
  
    
\bibitem{Dulaney:2010sq} 
  T.~R.~Dulaney and M.~I.~Gresham,
  Phys.\ Rev.\ D {\bf 81}, 103532 (2010)
  [arXiv:1001.2301 [astro-ph.CO]].
  
\bibitem{Gumrukcuoglu:2010yc} 
  A.~E.~Gumrukcuoglu, B.~Himmetoglu and M.~Peloso,
  Phys.\ Rev.\ D {\bf 81}, 063528 (2010)
  [arXiv:1001.4088 [astro-ph.CO]].
  
\bibitem{Watanabe:2010fh} 
  M.~-a.~Watanabe, S.~Kanno and J.~Soda,
  Prog.\ Theor.\ Phys.\  {\bf 123}, 1041 (2010)
  [arXiv:1003.0056 [astro-ph.CO]].
    
\bibitem{Hervik:2011xm} 
  S.~Hervik, D.~F.~Mota and M.~Thorsrud,
  JHEP {\bf 1111}, 146 (2011)
  [arXiv:1109.3456 [gr-qc]].
             
            
      
\bibitem{Barnaby:2012tk} 
  N.~Barnaby, R.~Namba and M.~Peloso,
  Phys.\ Rev.\ D {\bf 85}, 123523 (2012)
  [arXiv:1202.1469 [astro-ph.CO]].
  
                             
\bibitem{Martin:2007ue}
  J.~Martin and J.~'i.~Yokoyama,
  JCAP {\bf 0801} (2008) 025
  [arXiv:0711.4307 [astro-ph]].
              
\bibitem{ratra-pert}
  D.~Seery,
  JCAP {\bf 0908}, 018 (2009)
  [arXiv:0810.1617 [astro-ph]];
%
  R.~R.~Caldwell, L.~Motta and M.~Kamionkowski,
  Phys.\ Rev.\ D {\bf 84}, 123525 (2011)
  [arXiv:1109.4415 [astro-ph.CO]];
%
  C.~Bonvin, C.~Caprini and R.~Durrer,
  Phys.\ Rev.\ D {\bf 86}, 023519 (2012)
  [arXiv:1112.3901 [astro-ph.CO]];
%
  L.~Motta and R.~R.~Caldwell,
  Phys.\ Rev.\ D {\bf 85}, 103532 (2012)
  [arXiv:1203.1033 [astro-ph.CO]];
 %
  K.~Yamamoto,
  Phys.\ Rev.\ D {\bf 85}, 123504 (2012)
  [arXiv:1203.1071 [astro-ph.CO]];
%
  T.~Suyama and J.~'i.~Yokoyama,
  Phys.\ Rev.\ D {\bf 86}, 023512 (2012)
  [arXiv:1204.3976 [astro-ph.CO]];
%
  R.~K.~Jain and M.~S.~Sloth,
  arXiv:1207.4187 [astro-ph.CO];
  %
  M.~Shiraishi, S.~Saga and S.~Yokoyama,
  arXiv:1209.3384 [astro-ph.CO].


\bibitem{Linde:2007fr} 
  A.~D.~Linde,
  Lect.\ Notes Phys.\  {\bf 738}, 1 (2008)
  [arXiv:0705.0164 [hep-th]].



             
\bibitem{Kanno:2009ei} 
  S.~Kanno, J.~Soda and M.~-a.~Watanabe,
  JCAP {\bf 0912}, 009 (2009)
  [arXiv:0908.3509 [astro-ph.CO]].

\bibitem{Fujita:2012rb} 
  T.~Fujita and S.~Mukohyama,
  arXiv:1205.5031 [astro-ph.CO].


      
\bibitem{Watanabe:2010bu} 
  M.~-a.~Watanabe, S.~Kanno and J.~Soda,
  Mon.\ Not.\ Roy.\ Astron.\ Soc.\  {\bf 412}, L83 (2011)
  [arXiv:1011.3604 [astro-ph.CO]].

  
\bibitem{Pereira:2007yy} 
  T.~S.~Pereira, C.~Pitrou and J.~-P.~Uzan,
  JCAP {\bf 0709}, 006 (2007)
  [arXiv:0707.0736 [astro-ph]].
  
   
\bibitem{Gumrukcuoglu:2007bx} 
  A.~E.~Gumrukcuoglu, C.~R.~Contaldi and M.~Peloso,
  JCAP {\bf 0711}, 005 (2007)
  [arXiv:0707.4179 [astro-ph]].
  
\bibitem{Pitrou:2008gk} 
  C.~Pitrou, T.~S.~Pereira and J.~-P.~Uzan,
  JCAP {\bf 0804}, 004 (2008)
  [arXiv:0801.3596 [astro-ph]].
  

 
\bibitem{Dimopoulos:2008yv} 
  K.~Dimopoulos, M.~Karciauskas, D.~H.~Lyth and Y.~Rodriguez,
  JCAP {\bf 0905}, 013 (2009)
  [arXiv:0809.1055 [astro-ph]].
   
    
\bibitem{N1} 
  N.~Bartolo, E.~Dimastrogiovanni, S.~Matarrese and A.~Riotto,
  JCAP {\bf 0910}, 015 (2009)
  [arXiv:0906.4944 [astro-ph.CO]].
        
\bibitem{N2} 
  N.~Bartolo, E.~Dimastrogiovanni, S.~Matarrese and A.~Riotto,
  JCAP {\bf 0911}, 028 (2009)
  [arXiv:0909.5621 [astro-ph.CO]].
  
\bibitem{Manureview} 
  E.~Dimastrogiovanni, N.~Bartolo, S.~Matarrese and A.~Riotto,
  Adv.\ Astron.\  {\bf 2010}, 752670 (2010)
  [arXiv:1001.4049 [astro-ph.CO]].
       

    
\bibitem{Dey:2011mj} 
  A.~Dey and S.~Paban,
  JCAP {\bf 1204}, 039 (2012)
  [arXiv:1106.5840 [hep-th]].
    
\bibitem{N3} 
  N.~Bartolo, E.~Dimastrogiovanni, M.~Liguori, S.~Matarrese and A.~Riotto,
  JCAP {\bf 1201}, 029 (2012)
  [arXiv:1107.4304 [astro-ph.CO]].
         
        
\bibitem{Dey:2012qp} 
  A.~Dey and S.~Paban,
  arXiv:1205.2758 [astro-ph.CO].


\bibitem{Bartolo:2007ti} 
  N.~Bartolo, S.~Matarrese, M.~Pietroni, A.~Riotto and D.~Seery,
  JCAP {\bf 0801}, 015 (2008)
  [arXiv:0711.4263 [astro-ph]].
      
    
\bibitem{modulated}
  G.~Dvali, A.~Gruzinov and M.~Zaldarriaga,
  Phys.\ Rev.\ D {\bf 69}, 023505 (2004)
  [astro-ph/0303591];
%
  L.~Kofman,
  astro-ph/0303614;
  %
  F.~Bernardeau, L.~Kofman and J.~-P.~Uzan,
  Phys.\ Rev.\ D {\bf 70}, 083004 (2004)
  [astro-ph/0403315].


\bibitem{Matarrese:2003tk} 
  S.~Matarrese and A.~Riotto, 
  JCAP {\bf 0308}, 007 (2003) 
  [astro-ph/0306416]. 

    
\bibitem{triad}    
 C.~Armendariz-Picon,
  JCAP {\bf 0407}, 007 (2004)
  [astro-ph/0405267];
%
  A.~Maleknejad and M.~M.~Sheikh-Jabbari,
  arXiv:1102.1513 [hep-ph];
%
  P.~Adshead and M.~Wyman,
  Phys.\ Rev.\ Lett.\  {\bf 108}, 261302 (2012)
  [arXiv:1202.2366 [hep-th]];

 

    
    
\bibitem{pforms} 
  C.~Germani and A.~Kehagias,
  JCAP {\bf 0903}, 028 (2009)
  [arXiv:0902.3667 [astro-ph.CO]];
%
  T.~Kobayashi and S.~Yokoyama,
  JCAP {\bf 0905}, 004 (2009)
  [arXiv:0903.2769 [astro-ph.CO]];
 %
  T.~S.~Koivisto, D.~F.~Mota and C.~Pitrou,
  JHEP {\bf 0909}, 092 (2009)
  [arXiv:0903.4158 [astro-ph.CO]];
%
  T.~S.~Koivisto and N.~J.~Nunes,
  Phys.\ Lett.\ B {\bf 685}, 105 (2010)
  [arXiv:0907.3883 [astro-ph.CO]];
%
  C.~Germani and A.~Kehagias,
  JCAP {\bf 0911}, 005 (2009)
  [arXiv:0908.0001 [astro-ph.CO]].
%
  T.~S.~Koivisto and N.~J.~Nunes,
  Phys.\ Rev.\ D {\bf 80}, 103509 (2009)
  [arXiv:0908.0920 [astro-ph.CO]];
%
  T.~S.~Koivisto and F.~R.~Urban,
  Phys.\ Rev.\ D {\bf 85}, 083508 (2012)
  [arXiv:1112.1356 [astro-ph.CO]];
%
  A.~De Felice, K.~Karwan and P.~Wongjun,
  Phys.\ Rev.\ D {\bf 85}, 123545 (2012)
  [arXiv:1202.0896 [hep-ph]];
%
  F.~R.~Urban and T.~K.~Koivisto,
  JCAP {\bf 1209}, 025 (2012)
  [arXiv:1207.7328 [astro-ph.CO]];
%
  D.~J.~Mulryne, J.~Noller and N.~J.~Nunes,
  arXiv:1209.2156 [astro-ph.CO].






    
  


\end{thebibliography}
\end{document}